\documentclass[fleqn,usenatbib]{mnras}

\usepackage{newtxtext,newtxmath}
\usepackage[dvipsnames]{xcolor}
\usepackage[T1]{fontenc}
\usepackage [latin1]{inputenc}
\usepackage{multicol}
\usepackage{lscape}
\usepackage{rotating}

\DeclareRobustCommand{\VAN}[3]{#2}
\let\VANthebibliography\thebibliography
\AtBeginDocument{\def\thebibliography{\DeclareRobustCommand{\VAN}[3]{##3}\VANthebibliography}}

\usepackage{graphicx}	% Including figure files
\usepackage{amsmath}	% Advanced maths commands
\usepackage{hyperref}

	\usepackage{xspace}
	
	\newcommand{\eq}[1]{Eq.~(\ref{eq:#1})\xspace}
	\newcommand{\AU}{ \  \rm au}
	
	\newcommand{\Ms}{ \   \rm M_\odot }
	\newcommand{\Msyr}{ \ \rm M_\odot \,  \rm yr^{-1} }

	\newcommand{\Me}{ \  M_\oplus}

	\newcommand{\Omegak}{\Omega_{\rm K}}

	\newcommand{\alphag}{\alpha_{\rm g}}
	\newcommand{\alphat}{\alpha_{\rm t}}

\title[Modelling the giant planet occurrence rate]%{Modelling the giant planet occurrence rate from the Lick survey}
{Reproducing the stellar-mass dependence of the giant planet occurrence rate with pebble accretion models}

\author[H. F. Johnston et al.]{
Heather F. Johnston,$^{1, 2}$\thanks{E-mail: hf.johnston@outlook.com (UoE)}
O. Pani\'c,$^{2}$
S. Reffert,$^{3}$
B. Liu$^{4}$
and X. Ma,$^{1, 4}$
\\
$^{1}$ Department of Physics and Astronomy, University of Exeter, Stocker Road, Exeter EX4 4QL, UK\\
$^{2}$School of Physics and Astronomy , University of Leeds, Woodhouse, Leeds LS2 9JT, England\\
$^{3}$ Landessternwarte, Zentrum f\"{u}r Astronomie der Universit\"{a}t Heidelberg, K\"{o}nigstuhl 12, 69117 Heidelberg, Germany\\
$^{4}$Astronomical Institute, School of Physics, Zhejiang University, 38 Zheda Road, Hangzhou, 310027 China
}

\date{Accepted XXX. Received YYY; in original form ZZZ}

\pubyear{2026}

\begin{document}
\label{firstpage}
\pagerange{\pageref{firstpage}--\pageref{lastpage}}
\maketitle

% Abstract of the paper
\begin{abstract}
The stellar mass dependence of the unbiased giant planet occurrence rate may be the best statistical tool to constrain the formation of such planets.  This rate rises and falls as a function of stellar mass, peaking around stars of $\sim 1.7{-}2 \Ms$.  In this work, we carry out a population synthesis study, using pebble-driven core accretion model of planet formation, to investigate the planet formation conditions that may be responsible for this stellar-mass dependence.  We use the inferred giant planet occurrence rate of three combined homogenised radial velocity surveys (EXPRESS, PPPS, and Lick giant star survey) to constrain the models.  We find that we can produce a synthetic giant planet population with closely aligned occurrence and properties when we base our model on observationally-supported assumptions that accretion rates are higher and disk lifetimes are shorter around more massive stars.  We also find that in this scenario, the runaway gas accretion occurs at a larger orbital distance and earlier times as the stellar mass increases.

 %explore the times and locations at which runaway gas accretion occurs which can give us an indication of how giant planet composition might vary around host stars of different masses. 
%Giant planets are found most frequently around intermediate-mass stars ($\sim 1.7 \Ms$), a trend that has not yet been fully explained.  We carry out a planet population synthesis study to model the giant planet occurrence rate from the Lick survey by combining pebble-driven core accretion planet formation modelling with their observed giant planet population.  We find that x, y, and z. 
\end{abstract}

% Select between one and six entries from the list of approved keywords.
% Don't make up new ones.
\begin{keywords}
planets and satellites: formation -- stars: planetary systems -- methods: numerical
\end{keywords}

%%%%%%%%%%%%%%%%%%%%%%%%%%%%%%%%%%%%%%%%%%%%%%%%%%

%%%%%%%%%%%%%%%%% BODY OF PAPER %%%%%%%%%%%%%%%%%%

\section{Introduction}

Giant planets - while being only observed around $10{-}20 \%$ of stars \citep{Cumming2008ThePlanets, Johnson2010GIANTPLANE, Mayor2011ThePlanets} - play a vital role in shaping the architecture of planetary systems that host them.  Studying the conditions of their formation is pivotal in furthering our understanding of dynamics and habitability of planetary systems, including our own Solar System \citep{Liu2022EarlyDisk}. 

The occurrence rate of giant planets $\eta_{\rm J}$ is found to increase with stellar metallicity across both main-sequence and evolved stars \citep{Gonzalez1997TheConnection, Santos2001ThePlanets, Santos2004SpectroscopicFormation, Fischer2005THE1, Udry2007StatisticalExoplanets, Johnson2010GIANTPLANE, Sousa2011SpectroscopicCorrelation, Mortier2012TheStars, Mortier2013OnCorrelation, Reffert2015PreciseMetallicity, Wolthoff2022PreciseSurveys, Pan2025DependenceModel}.  \cite{Ghezzi2018RetiredMass} propose that the probability of giant planet formation is around one-to-one for the total amount of metals in the protoplanetary disk.  Their sample of evolved A-stars, along with FGKM dwarfs, confirms that $\eta_{\rm J}$ increases with both metallicity and stellar mass up to $2 \Ms$ \citep{Ghezzi2018RetiredMass}.  

However, while radial velocity (RV) surveys have surmised that $\eta_{\rm J}$ increases with stellar mass between $0.5{-}2 \Ms$ \citep{Johnson2010GIANTPLANE, Bowler2010RetiredStars, Borgniet2019ConstraintsImaging} - this does not hold for stars more massive than $2 \Ms$.  \cite{Reffert2015PreciseMetallicity} and \cite{Jones2015Giant67851c} both reported a likely peak in $\eta_{\rm J}$ at $1.5{-}2 \Ms$ that drops quickly for more massive stars.  Giant planets are incredibly rare around stars more massive than $2.5 \Ms$ ($\eta_{\rm J} \sim 0$).  Additionally, in \cite{Wolthoff2022PreciseSurveys}, they combined data from three different RV surveys (EXPRESS, PPPS, and Lick giant star survey) placed the peak of $\eta_{\rm J}$ around stars of $1.7 \Ms$.  Direct imaging surveys have also inferred that the frequency of giant planets increases with stellar mass \citep{Wagner2022TheStars}.  

The exact origin of the stellar mass dependence of $\eta_{\rm J}$ rate is still unclear.  \cite{Kennedy2009StellarDispersal} were among the first to discuss possible physical reasons for the paucity of close-in giant planets around massive stars, noting faster disk dispersal and shorter disk lifetimes act to suppress giant planet formation around more massive stars.  This disk-lifetime mechanism has since been invoked by \cite{Burkert2007THEM, Currie2009OnStars, Alibert2011ExtrasolarMasses} to explain the paucity of close-in giant planets around intermediate-mass stars. The lifetimes of protoplanetary disks, driven by pre-main sequence (PMS) stellar evolution, are thought to have major influence over the success of giant planet formation - and resulting $\eta_{\rm J}$ rate. 

Testing this has been challenging because, until recently, there was no complete, unbiased sample of young intermediate-mass stars with which to measure disc lifetimes; earlier work relied largely on Herbig Ae/Be stars, which are older and not representative. The X-shooter survey of \cite{Iglesias2023X-ShooterEvolution} is the most complete unbiased sample to date within 300 pc. The vast majority of its intermediate-mass stars host debris-like or transitional disks rather than protoplanetary ones, and their inner disks disperse differently from those of low-mass stars, suggesting shorter disc lifetimes. \cite{Yasui2014RapidStars} measured the disk lifetime around intermediate-mass stars ($1.5{-}7$ M$_{\odot}$) and found that the innermost dust disk ($\sim 0.3$ au) disperses on a timescale of $\sim 3$ Myr, while the inner disk ($\sim 5$ au) persists for $\sim 6.5$ Myr. They suggest this rapid early disappearance of the innermost disk could be a major reason for the paucity of close-in planets around intermediate-mass stars.

A second, related effect is the evolution of the host star itself.  The luminosity evolutionary tracks of PMS stars with $M_{\star} \geq 1.5 \Ms$ differ significantly from their solar-mass counterparts, even at young ages.  Low-mass stars decrease in luminosity as they evolve towards the main sequence.  In contrast, stars with masses $\geq 1.5$ M$_{\odot}$ increase in luminosity as they approach the main sequence, warming their disks within a few Myr \citep{Miley2020TheDiscs}. This elevated disk temperature increases the gas scale height, reducing pebble accretion efficiency and slowing planet formation \citep{Johnston2025TheFormation}.  
%They deviate from the well-constrained X-ray photoevaporation rate for far-ultraviolet (FUV) \citep{Kunitomo2021PhotoevaporativeStars}.  This shift in photoevaporation mechanism could explain the uncommonly low extreme ultraviolet (EUV) and X-ray luminosities found in grain-depleted protoplanetary disks around intermediate-mass stars \citep{Nakatani2021PhotoevaporationRemnants}.  Due to absorption by the stellar atmosphere and the convective layer \textbf{of the star becoming radiative}, mass loss rates were diminished which resulted in longer lifetimes of gas disks around A-type stars.     

In this paper, we use the pebble-driven core accretion planet formation model from \cite{Liu2019Super-EarthMasses} and \cite{Johnston2024FormationStars} to reproduce the observed planet occurrence rates as a function of mass and metallicity, derived in the studies by \cite{Reffert2015PreciseMetallicity, Wolthoff2022PreciseSurveys}, through planet population synthesis. For each stellar mass-metallicity combination in the RV observed sample of \cite{Wolthoff2022PreciseSurveys}, we conduct 100 numerical simulations of the growth of single embryos.
%In the simulation, we employ stellar-mass related parameter sampling of initial accretion rate $\dot{M}_{\rm g0}$;  and birth mass of the embryo $M_{\rm p0}$.  We allow both birth location of the embryo $r_{0}$ and initial disk size $R_{\rm d0}$ to be randomly sampled within reasonable physical boundaries.  \textbf{We determine the mass of the planetary embryo through streaming instability.}   We fix the birth time of the embryo to be $t_{0} = 0.1$ Myr for reasons stated above.

 %While we solely employ X-ray driven photoevaporation in our model for our stellar mass range of $0.9{-}5 \Ms$, we only allow embryos to be born at times $\leq 0.1$ Myr.  %Furthermore, we do consider the impact of PMS stellar evolution.  
 %The implications of this are discussed in more detail in Section \ref{sec:pms_ev}. 

The aim of this work is to reproduce the giant planet occurrence rate from the observations via planet population synthesis using our model and establish the reasons behind the distribution of the giant planet population \citep{Reffert2015PreciseMetallicity, Wolthoff2022PreciseSurveys}.  In order to do this, we seek to identify the key physical properties that drive the rise and fall of the giant planet occurrence rate as a function of stellar mass.  The paper is organised as follows.  The key principles of the model are described in Section \ref{sec:method}.  In Section \ref{sec:setup}, we recount the setup of the model and our relevant assumptions.  Section \ref{sec:results} discusses the results of our population synthesis study and compare to the giant planet occurrence rate as functions of stellar mass and metallicity.  Penultimately, we discuss where and when our giant planets accrete the bulk of their mass and how the orbital distances of our synthetic giant planet population compare to the observed distances in Section \ref{sec:prop}.  %Important physical parameters of both population synthesis studies and how this fits into ongoing observational findings are discussed in Section \ref{sec:discussion}.  
In Section \ref{sec:conc}, our final conclusions are drawn and summarised.

\section{Method}
\label{sec:method}
We employ the planet formation model of \cite{Liu2019Super-EarthMasses}, originally adapted in \cite{Johnston2024FormationStars},  and summarise the key equations in this section.  Readers are recommended to  go through their Section 2 for details.  The disk condition presented here are adapted to host stars of masses $0.9{-}5 \Ms$, with the addition of evolving stellar luminosity, in order to encompass the stellar mass range from the observed RV sample in \cite{Wolthoff2022PreciseSurveys}.   %intermediate-mass star of $1{-}2.4 \Ms$. 

\subsection{Disk model}
\label{sec:disk}

We adopt a conventional $\alpha$ viscosity prescription for the gas disk angular momentum transport \citep{Shakura1973BlackAppearance}.  The disk accretion and surface density are linked through $\dot M_{\rm g}{=}3 \pi \nu \Sigma_{\rm g}$ with viscosity $\nu{=}\alphag c_{\rm s} H_{\rm g}$, where  $c_{\rm s}{=}H_{\rm g} \Omegak$ is the sound speed, $H_{\rm g}$ is the gas disk scale height and $\Omegak$ is the Keplerian frequency. The dimensionless viscous coefficient $\alphag$ determines the global disk evolution and gas surface density.  We set it to be a fixed value of $10^{-2}$, the same as \cite{Liu2019Super-EarthMasses, Johnston2024FormationStars}.  %\textcolor{red}{comment on observational constraints}

We note that the viscous disk framework with a high $\alpha_{\textrm{g}}=10^{-2}$ is a strong assumption. Protoplanetary disks may be laminar and magnetic braking may dominate angular momentum transport, instead of viscosity (\textit{e.g.} \cite{Bai2013Wind-drivenWind}).  The surface density profile can be qualitatively different \citep{Suzuki2016EvolutionWinds}, leading to slowing down the migration speed of protoplanets \citep{Ogihara2018FormationWinds}.  The temperature (and thus scale height) structure can also be different from the classical picture (\textit{e.g.} \cite{Mori2025Long-termComposition}).  These limitations should be kept in mind when interpreting our results. 

We also note that the origin of the $\dot{M}_{\textrm{g}}-M_{\star}$ relation remains an open question. It may reflect a stellar-mass dependence of the angular momentum transport efficiency $\alpha_{\textrm{g}}$, or arise from differences in the photoevaporation rate as a function of stellar mass \citep{Ercolano2014TheDiscs}.  If $\alpha_{\textrm{g}}$ depends on $M_{\star}$, both the disk temperature structure and viscous timescale would be modified, which in turn would affect the planet formation outcome. Exploring these possibilities is beyond the scope of the present work but represents an important direction for future modelling.  

The evolution of the disk accretion rate is expressed as
  \begin{equation}
 \dot{M}_{\rm g} = \begin{cases}
 {\displaystyle   \dot{M}_{\rm g0} \left[ 1 + \frac{t}{\tau_{\rm vis}} \right] ^{-\gamma} }, 
 \hfill  \ t{<}t_{\rm pho}, \\
 {\displaystyle \dot{M}_{\rm g0} \left[ 1 + \frac{t}{\tau_{\rm vis}} \right]^{-\gamma} \exp \left[ - \frac{t - t_{\rm pho}}{\tau_{\rm pho}} \right], }  \hfill \   t{\geq}t_{\rm pho},\\
\end{cases}
\label{eq:mdotg}
\end{equation}
where  $\dot M_{\rm g0}$ is the initial disk accretion rate; $t_{\rm pho}$ is the time at which $\dot{M}_{\textrm{g}}$ decreases to $\dot{M}_{\textrm{pho}}$; and the $t-t_{\textrm{pho}}$ term refers to the mass loss due to stellar photoevaporation; $\tau_{\rm vis}$ and $\tau_{\rm pho}$ are the characteristic viscous accretion and photoevaporation timescales from Eqs. (3) and (7) of \cite{Liu2019Super-EarthMasses}, $\gamma =( 5/2+s) / (2+s)$, and $s$ is the gas surface density gradient.

We note that this simple exponential decay model may not fully capture the complexity of disk evolution. In particular, the decay timescale in \citep{Owen2012OnPhotoevaporation} is determined by the viscous timescale at the gap location \citep{Clarke2001TheSwitch}, rather than the ratio of disk mass to accretion rate.  A slower decay would prolong the availability of gas for planet formation, which could increase the giant planet occurrence rates. This represents a limitation of our model and should be kept in mind when interpreting our results. 

We also note that our implementation of photoevaporation does not include inside-out clearing, in which the disk is first cleared from the inside out by photoevaporation and then the outer disk draings (\textit{e.g.} \cite{Burkert2007THEM}).  If inside-out clearing were included, giant planets might be halted at larger orbital radii rather than migration all the way to the inner disk edge. This represents a limitation of our model and could affect the predicted distribution of close-in giant planets.  

 We specifically account for the X-ray driven photoevaporation, and the critical mass-loss rate is given by \citep{Owen2012OnPhotoevaporation}:
 
 \begin{equation}
 \begin{split}
   \frac{ \dot{M}_{\rm pho}}{ M_\odot \rm yr^{-1}}  = 6 \times 10^{-9} \left( \frac{M_{\star}}{1 \ M_\odot} \right)^{-0.07} \left( \frac{L_{X}}{10^{30} \rm \ ergs^{-1}} \right)^{1.14}  \\
    {=} 3 \times 10^{-8} \left( \frac{M_{\star}}{2.4 \ M_\odot} \right)^{1.6},
    \end{split}
    \label{eq:pho}
  \end{equation}

where we have substituted the $L_{X}{-}M_{\star}$ relation from \cite{Owen2012PlanetaryHydrodynamics} to obtain the second line.

%It is important to note that \cite{Kunitomo2021PhotoevaporativeStars} examined long-term disk evolution of stars in the range of 0.5-5 $\textrm{M}_{\odot}$.  They found that the X-ray luminosity of more massive stars decreases in just 1 Myr while the FUV luminosity rapidly increases.  The critical mass for this FUV-dominated photoevaporative mass-loss regime is thought occur at $2.5 \  \textrm{M}_{\odot}$, when the Kelvin-Helmholtz timescale is comparable with the disk dispersal timescale \citep{Kunitomo2021PhotoevaporativeStars}.  Thus, for the majority of our models, X-ray photoevaporation as the dominant mass-loss mechanism post-viscous evolution is a valid assumption and we discuss the implications of the shift in photoevaporative mechanism for the most massive stars in our sample, \textit{e.g.} $M_{\star} > 2.5$ M$_{\odot}$.  

We employ a two-component disk structure, including an inner viscously heated region and an outer stellar irradiated region. The gas surface density and disk aspect ratio are adopted from Eqs. (8)-(13) of \cite{Liu2019Super-EarthMasses}, where $[\rm vis]$ refers to the inner viscously heated disk and $[\rm irr]$ refers to the outer disk region heated entirely by stellar irradiation (where the disk is optically thin):
  \begin{equation}
 \frac{\Sigma_{\rm g}}{ \rm g \ cm^{-2}}  =
  \begin{cases}
 {\displaystyle   466
    \left( \frac{\dot M_{\rm g}}{10^{-7} \Msyr} \right)^{1/2}  \left(\frac{M_{\star}}{2.4 \ M_{\odot}} \right)^{1/8} \left( \frac{\alpha_{\textrm{g}}}{10^{-2}} \right)^{-3/4} }
    \\
   {\displaystyle  \left( \frac{\kappa_{0}}{10^{-2}} \right)^{-1/4} \left(\frac{r}{1 \AU} \right)^{-3/8}  }
     \hfill  [\mbox{vis}],  \vspace{0.6cm}\\
 {\displaystyle  1613 \left( \frac{\dot M_{\rm g}}{10^{-7}  \Msyr} \right)
\left(\frac{M_{\star}}{2.4 \ M_{\odot}} \right)^{9/14} \left( \frac{\alpha_{\textrm{g}}}{10^{-2}} \right)^{-1}} \\
{\displaystyle  \left(\frac{L_{\star}}{33 \ L_{\odot}} \right)^{-2/7}  \left(\frac{r}{1 \AU} \right)^{-15/14} }
  \hfill  [\mbox{irr}], 
\end{cases}
\label{eq:sigma}
\end{equation}
  \begin{equation}
h_{\rm g} = \begin{cases}
 {\displaystyle   0.043
\left( \frac{\dot M_{\rm g}}{10^{-7} \Msyr} \right)^{1/4}
\left(\frac{M_{\star}}{2.4 \ M_{\odot}} \right)^{-5/16} \left( \frac{\alpha_{\textrm{g}}}{10^{-2}} \right)^{-1/8}} \\
   {\displaystyle  \left( \frac{\kappa_{0}}{10^{-2}} \right)^{1/8} \left(\frac{r}{1 \AU} \right)^{-1/16}  }  
     \hfill  [\mbox{vis}],  \vspace{0.6cm}\\
 {\displaystyle   0.025
    \left(\frac{M_{\star}}{2.4 \ M_{\odot}} \right)^{-4/7} 
    \left(\frac{L_{\star}}{33\ L_{\odot}} \right)^{1/7}  }\\
     {\displaystyle    \left(\frac{r}{1 \AU} \right)^{2/7} } 
       \hfill  [\mbox{irr}].
\end{cases}
\label{eq:aspect}
\end{equation}

The temperature profile of the disk follows:

 \begin{equation}
 \frac{T_{\rm g}}{ \rm K}  =
  \begin{cases}
 {\displaystyle   280
    \left( \frac{\dot M_{\rm g}}{10^{-7} \Msyr} \right)^{1/2}  \left(\frac{M_{\star}}{2.4 \ M_{\odot}} \right)^{3/8} } \\
   {\displaystyle  \left( \frac{\alpha_{\rm g}}{10^{-2}} \right)^{-1/4} \left( \frac{\kappa_{0}}{10^{-2}} \right)^{-1/4}  \left(\frac{r}{1 \AU} \right)^{-9/8}  }
     \hfill  [\mbox{vis}],  \vspace{0.6cm}\\
 {\displaystyle  236 
\left(\frac{M_{\star}}{2.4 \ M_{\odot}} \right)^{-1/7} \left(\frac{L_{\star}}{2.4 \ L_{\odot}} \right)^{2/7}} \\
{\displaystyle   \left(\frac{r}{1 \AU} \right)^{-1/2} \phi_{\rm inc}^{1/4} }
  \hfill  [\mbox{irr}], 
\end{cases}
\label{eq:temp}
\end{equation}

where $M_{\star}$, $L_{\star}$ are stellar mass and luminosity,  and $r$ is the radial distance to the central star. 
The transition radius between these two disk regions is written as
\begin{equation}
    \begin{split}
   r_{\rm tran} =  4.9
    \left( \frac{\dot M_{\rm g}}{10^{-7}  \Msyr} \right)^{28/39}
    \left(\frac{M_{\star}}{2.4 \ M_{\odot}} \right)^{29/39} \\
    \left(\frac{L_{\star}}{2.4 \ L_{\odot}}\ \right)^{-16/39}    \AU.
     \label{eq:rtrans}
    \end{split}
\end{equation}
 
 The main disk solid reservoir - pebbles - are assumed to be constituted of $35\%$ water ice and $65\%$ silicate \citep{Liu2019Super-EarthMasses}.  The water-ice line $r_{\rm H_2O}$ is derived by equating the saturated pressure and $\rm H_2 O$ vapor pressure (approximately in the disk radius at the temperature of $170$ K).
 
We do not account for any detailed grain growth processes. Instead, we  assume that the dust has already fully grown and treat their final size to be a free parameter, indicative as either a fixed physical size or a fixed Stokes number. The above simplification ideally mimics the grain growth is limited by the bouncing or fragmentation barriers \citep{Guttler2010TheBarrier}.

The inner edge of the disk is truncated by the stellar magnetospheric torque \citep{Lin1996OrbitalLocation}.  We approximate the co-rotation cavity radius of the host star %to the magnetospheric cavity radius 
from \cite{Mulders2015ARates} as:

\begin{equation}
\label{eq:r_in}
    r_{\rm in} {=} \sqrt [3] {\frac{GM_{\star}}{ \Omega_{\star}^{2}}} {\simeq} 0.06 \left( \frac{M_{\star}} {2.4 M_{\odot}} \right)^{1/3} \rm au,
\end{equation}
where $\Omega_{\star}$ is the stellar spin frequency.  

In our model, the inner disk edge acts as the point at which inward migration and solid accretion are halted. This cutoff location from \cite{Mulders2015ARates} is in the range of $\sim 0.05{-}0.1\ \rm au$, depending on stellar mass, and is in agreement with the recent models from \cite{Flock2019PlanetDisks} for FGK and M-type stars.

We note that the surface density evolution is not solved self-consistently in our model. Instead, we adopt an analytic prescription for the disk accretion rate (Equation \ref{eq:mdotg}) and surface density profile (Equation \ref{eq:sigma}), following \cite{Liu2019Super-EarthMasses}.  Population synthesis models that solve the surface density evolution, such as the Bern model (\textit{e.g.} \cite{Emsenhuber2021TheResults}), may produce different disk structures and hence different planet formation outcomes. In particular, a gap-opening by photoevaporation at around a few au, which is not included in this work, could affect the final location of giant planets \cite{Alexander2012DesertsClearing}. This represents a limitation of our model and motivates future work. 

\subsection{Planet growth and migration}
\label{sec:growth}

Core mass growth proceeds via pebble accretion. Embryo formation is treated through streaming instability, a powerful mechanism that forms planetesimals from the collapse of many mm-sized pebbles \citep{Youdin2005STREAMINGDISKS}.  It is a method that is particularly effective at forming planetesimals \citep{Johansen2007RapidDisks, Johansen2012AddingInstabilities, Bai2010DynamicsFormation, Simon2016THESELF-GRAVITY, Schafer2017InitialInstability, Abod2019TheGradient, Li2019DemographicsInstability}. 
 \cite{Liu2020Pebble-drivenDwarfs} summarised from the literature that streaming instability planetesimal formation simulations show the birth masses of embryos depend on their evolutionary time and disk locations, which can be expressed as:

\begin{equation}
    \frac{M_{\rm p0}}{M_{\oplus}} = 6 \times 10^{-2} \left( \gamma \pi \right) ^{1.5}  
    \left (\frac{h_{\rm g}}{0.05} \right ) ^{3} \left ( \frac{M_{\star}}{2.4\ \rm M_{\odot}} \right).
    \label{eq:embryo_mass}
\end{equation}

The self-gravity parameter $\gamma$ that quantifies the relative strength between the gravity and tidal shear is given by  
\begin{equation}
    \gamma \equiv \frac{4 \pi G \rho_{\rm g}}{\Omega^{2}_{\rm K}}.
\end{equation}

The solid accretion rate onto the planet's core reads 
\begin{equation}
\label{eq:peb_acc}
    \dot{M}_{\rm PA} = \epsilon_{\rm PA}\dot{M}_{\rm peb} = \epsilon_{\rm PA} \xi \dot{M}_{\rm g},
\end{equation}
where $\epsilon_{\rm PA}$ is the efficiency of pebble accretion, the formulas of which are adopted from \citet{Liu2018CatchingPlanets} and \cite{Ormel2018CatchingPebbles} that include both $2$D and $3$D regimes and expressed as:

\begin{equation}
\label{eq:peb_acc_eff}
    \epsilon_{\rm PA} = \sqrt{\epsilon_{\rm PA, 3D}^{2} + \epsilon_{\rm PA, 2D}^{2}},
\end{equation}
where $\epsilon_{\rm PA, 2D}$ and $\epsilon_{\rm PA, 3D}$ are defined from \cite{Liu2018CatchingPlanets} and \cite{Ormel2018CatchingPebbles}, respectively.

The pebble scale height is given by:
 \begin{equation}
 \label{eq:pebh}
    H_{\rm peb} = \sqrt{\frac{\alphat}{\alphat + \tau_{\rm s}}} H_{\rm g},
\end{equation}
where $\tau_{\rm s}$ is the Stokes number of the particles and $\alpha_{\rm t}$ is the turbulent diffusion coefficient, approximately equivalent to the midplane turbulent viscosity when the disk is driven by magnetorotational instability \citep{Johansen2005DUSTTURBULENCE, Zhu2015DustDiffusion, Yang2017ConcentratingInstability}. 

Whether pebble accretion is in $2$D or $3$D depends on the ratio between the radius of pebble accretion and the vertical layer of pebbles \citep{Morbidelli2015TheCores}.

Similar to \cite{Liu2019Super-EarthMasses,Liu2020Pebble-drivenDwarfs}, we assume that the pebble and gas flux ratio remains a constant such that $\xi{=}\dot{M}_{\rm peb}/\dot{M}_{\rm g}$. 
In the early disk phase, $\xi = \xi_{0}$ is spatially constant and equal to the initial disk metallicity.  While pebbles lose approximately 35\% of their mass across the water-ice line due to sublimation, we neglect this size change and its effect on the pebble flux, so $\xi$ remains spatially constant at the ice line.
%When pebbles have a low Stokes number, they are well-coupled to the disk gas and drift inwards at the same rate, such that the pebble-to-gas surface density ratio reflects the initial disk metallicity.  When the Stokes number is high, pebbles drift faster than the gas, and in order to maintain a constant flux ratio, $\Sigma_{\rm peb}/\Sigma_{\rm g}$ falls below the nominal disk metallicity. 
We also neglect the metallicity enrichment in the late rapid gas removal phase when the stellar photoevaporation dominates.

A growing planet begins to clear the surrounding gas and open a partial gaseous annular gap, causing the inward pebble flux to be halted at the outer edge of the planetary gap and solid accretion to terminate \citep{Lambrechts2014SeparatingAccretion}.  This planetary mass threshold is referred to as the pebble isolation mass, and we adopt the formula from \cite{Bitsch2018Pebble-isolationGiants}:

 \begin{equation}
   \begin{split}
 M_{\rm iso} = & 16 \left( \frac{h_{\rm g}}{0.03} \right)^3 \left( \frac{M_{\star}}{3 \ M_{\odot}} \right) \left[ 0.34 \left(  \frac{ -3}{ {\rm log_{10}} \alphat}  \right)^4 + 0.66 \right] \\
  & \left[ 1-   \frac{ \partial  {\rm ln } P /  \partial {\rm ln} r +2.5} {6}        \right] M_{\oplus}.
 \label{eq:m_iso}
 \end{split}
 \end{equation}
 
We note that a non-negligible fraction of pebble fragments at the above-mentioned planetary gap can still pass through and replenish the planetary envelope \citep{Chen2020ThePhase} as well as the inner disk region \citep{Liu2022NaturalDisk, MarkusStammler2023LeakyPlanets}. We do not model this effect here.

When the low-mass planets still accrete pebbles, the solid materials entering the planetary atmosphere would generate sufficient heating to prevent the further contraction of the surrounding gas. Therefore,  we only follow the gas accretion when the planet grows beyond $M_{\rm iso}$. The gas accretion rate can be expressed as 
\begin{equation}
    \dot{M}_{\rm p, g} = \mathrm{min} \left[ \left( \frac{\mathrm{d}M_{\rm p, g}}{\mathrm{dt}} \right)_{\rm KH} , \left( \frac{\mathrm{d}M_{\rm p,g}}{\mathrm{dt}} \right)_{\rm Hill} , \dot{M}_{\rm g} \right].
\end{equation}

We adopt the gas accretion rate based on \cite{Ikoma2000FORMATIONOPACITY}:

\begin{equation}
\label{eq:kh}
    \left( \frac{\mathrm{d}M_{\rm p, g}}{\mathrm{dt}} \right)_{\rm KH} = 10^{-5} \left( \frac{M_{\rm p}}{10\ \rm M_{\oplus}} \right)^{4} \left( \frac{\kappa_{\rm env}}{1\ \rm cm^{2}\ \rm g^{-1}} \right ) ^{-1} \rm M_{\oplus} yr^{-1}
\end{equation}

where $\kappa_{\rm env}$ is the planet's gas envelope opacity.  We assume $\kappa_{\rm env}{=} 0.05\ \rm cm^{2}\ g^{-1}$ and that it does not vary with metallicity, as in \cite{Liu2019Super-EarthMasses}.

Physically, the Kelvin-Helmholtz thermal contraction in Equation \ref{eq:kh} promotes the envelope mass growth in the first place.  As the planet grows, the accreted gas is further limited by the total amount entering the planetary Hill Sphere, adopted from \cite{Liu2019Super-EarthMasses} below:

\begin{equation}
    \left( \frac{\mathrm{d}M_{\rm p,g}}{\mathrm{dt}} \right)_{\rm Hill} {=} f_{\rm acc} \nu_{\rm H} R_{\rm H} \Sigma_{\rm Hill} = \frac{f_{\rm acc}}{3\pi} \left( \frac{R_{\rm H}}{H_{\rm g}} \right)^{2} \frac{\dot{M}_{\rm g}}{\alpha_{\rm g}} \frac{\Sigma_{\rm gap}}{\Sigma_{\rm g}}. 
\end{equation}

where $\nu_{\rm H}{=} R_{\rm H} \Omega_{\rm k}$ is the Hill velocity; $R_{\rm H}{=}(M_{\rm p} / 3\ \rm M_{\star})^{1/3}$ is the Hill radius of the planet; and $\Sigma_{\rm Hill}$ is the gas surface density near the planet's Hill sphere.  In this paper, we set $f_{\rm acc}{=}0.5$ to be the gas fraction that can be accreted by the planet's Hill sphere.  Once the planet becomes  sufficiently massive, accretion is dictated by the global disk gas inflow across the planetary orbit.

We adopt a combined migration formula by incorporating both type I and type II regimes \citep{Kanagawa2018RadialPlanet}: 
\begin{equation}
  \dot{r} =  f_{\rm tot} \left( \frac{M_{\rm p}}{M_{\star}} \right) \left( \frac{\Sigma_{\rm g} r^{2}}{M_{\star}} \right) h_{\rm g}^{-2} v_{\rm K},
  \label{eq:mig}
\end{equation}
and the migration coefficient is given by
 \begin{equation}
f_{\rm tot} = f_{\rm I} f_{\rm s} + f_{\rm II} (1- f_{\rm s}) \frac{1} {\left(\frac{M_{\rm p}}{M_{\rm gap}} \right)^2},
\end{equation}
where $M_{\rm p}$ is the mass of the planet, $v_{\rm K}$ is the Keplerian velocity and $f_{\rm I}$ and  $f_{\rm II}$ correspond to the type I and II migration prefactors, respectively. The type II migration coefficient $f_{\rm II}{=}-1$ whereas the type I migration coefficient $f_{\rm I}$ sets the direction and strength of the type I torque, determined by the disk thermal structure and local turbulent viscosity $\alpha_{\rm t}$ \citep{Paardekooper2011ADiffusion}.  
 We choose the smooth function $f_{\rm s}^{-1}{=}1+ (M_{\rm p}/M_{\rm gap})^{4}$. This ensures that $\dot{r}{\simeq } \dot{r}_{\rm I}$ when $M_{\rm p}{\ll} M_{\rm gap}$ and $\dot{r}{\simeq } \dot{r}_{\rm I}/(M_{\rm p}/M_{\rm gap})^{2}$ when $M_{\rm p}{\gg} M_{\rm gap}$ \citep{Kanagawa2018RadialPlanet}.

\subsection{Stellar luminosity evolution} 
\label{sec:pms_ev}  

As a wholly convective star evolves along its' Hayashi track until it develops a radiative core, the stellar luminosity evolves in response.  The parameterization of the stellar luminosity evolution depends on stellar mass.  Importantly, $L_{\star}$ decreases with age for low-mass stars ($M_{\star} \leq 1.5\ \Ms$), but has an upward turn and remains high for more massive stars ($M_{\star} {>} 1.5\ \Ms$) due to their transition to the radiative regime \citep{Palla1993TheStars, Siess2000AnStars, Baraffe2002EvolutionaryAges}.  Furthermore, \cite{Johnston2025TheFormation} found that giant planet formation is sensitive to the evolution of stellar luminosity, specifically the locations and times at which giant planet formation can occur depend on it. High stellar luminosity causes an increase in the scale height of the gas, which may decrease the efficiency of pebble accretion, making it more challenging to form giant planets \citep{Johnston2025TheFormation}.

%\begin{figure}
%\centering
%\includegraphics[width = \columnwidth]%{disk_size.png}
%{fig1_miley_pms-hr.png}
%\caption{Results of the stellar evolution models from \citep{Miley2020TheDiscs}. The dotted lines denote isochrones for specified timesteps.}
%\label{fig:pms}
%\end{figure}

%In \cite{Miley2020TheDiscs}, their inclusion of stellar luminosity evolution resulted in temperature profiles of disks around low- and intermediate-mass stars that begin to diverge at around $2\ \rm Myr$.  Disks around stars where $M_{\star} {\geq} 1.5\ \rm M_{\odot}$ become warmer over time due to the increasing stellar luminosity, while disks around stars where $M_{\star} {\leq 1.5} \ \rm M_{\odot}$ cool in temperature as the stellar luminosity decreases \citep{Miley2020TheDiscs}.

Thus in our model, we now adopt an evolving stellar luminosity derived from the evolutionary tracks in the Siess models \citep{Siess2000AnStars}.   We select stellar luminosity values at timesteps $0.1{-}10\ \rm Myr$ with solar metallicity ($Z{=}0.01$) for stellar masses in the range of $0.9{-}5\ \Ms$ accounting for the majority of the stellar masses in the observational sample from \cite{Wolthoff2022PreciseSurveys} and interpolate the values in between. 

We recognise that our model is limited by the sole X-ray photoevaporation prescription.  This is due to X-ray rates being well-constrained whereas FUV photoevaporation rates are notoriously challenging to quantify due to FUV interacting with the dust.  %its' intrinsic dependence on dust distributions within the disk.  
This regime switch typically occurs when stars become radiative \citep{Kunitomo2021PhotoevaporativeStars}.   %in stars more massive than $2.5\ \rm M_{\odot}$ as they differ significantly from their solar-mass analogs.  
Intermediate-mass stars become radiative early on in terms of disk lifetime $(1{-}3\ \rm Myr)$, causing their X-ray rates to drop and FUV to become the dominant mass-loss mechanism \cite{Kunitomo2021PhotoevaporativeStars}.  While this is only a main concern for sampled stars in the range of $2.5{-}5\ \Ms$, where the observed giant planet occurrence rate remains incredibly low ($<5\%$) \citep{Reffert2015PreciseMetallicity, Wolthoff2022PreciseSurveys}, it is still important to consider in the context of our model.  Our attempt to mitigate this is by having all our model embryos born at time $0.1\ \rm Myr$ - when X-rays are still the dominant photoevaporation mechanism for most of our stellar mass range.  %We discuss the ramifications of this in Section \ref{sec:discussion}.  

\section{Model assumptions}
\label{sec:setup}
We carry out a population synthesis study, resulting in the final masses and locations of planets formed in our models.  This allows us to explore the frequency and efficiency of planet formation with varying initial disk conditions, and to investigate the birth locations of embryos that grow into successful giant planets.  % corresponding to the resulting planet orbits.  
The disk model is as described in Section \ref{sec:disk}. We proceed by conducting $100$ numerical simulations for each stellar mass and metallicity combination from the \cite{Wolthoff2022PreciseSurveys} sample for the growth of a single embryo by Monte Carlo sampling each of the initial conditions:  $ R_{\rm d0}$ sampled between 20 and 200~AU, $ r_0$ sampled logarithmically between 0.1 and 100~AU.  The birth time is fixed at $t_{0}{=}0.1\ \rm Myr$ and the pebble size $R_{\rm peb}{=}1\ \rm mm$.  The initial birth mass of the embryos is derived from Equation \ref{eq:embryo_mass}.  We use the stellar masses and metallicities from the homogenised stellar sample in \cite{Wolthoff2022PreciseSurveys} (described below).  The adopted distributions of disk and stellar parameters are summarised in Table \ref{tab:pps}.  These parameters are discussed in more detail below.

%\subsubsection{Disk mass}

%Gas disk mass from can be expressed as \citep{Liu2019Super-EarthMasses}: 
%\begin{equation}
%\label{eq:gmass}
%    M_{\rm d}{=} \int^{r_{\rm d}}_{0} 2 \pi r \Sigma_{\rm g} (r, t) dr.
%\end{equation}

%This was the assumption used to achieve the first set of giant planet occurrence rate calculations.

%We can then manipulate $M_{\rm d}$ by varying the gas surface density, $\Sigma_{\rm g}$.  There has been much theoretical conjecture that having a more massive disk is most likely to form more massive planets, due to the readily available abundance of material.  In order to explore this in the purely core accretion regime, we allow $M_{\rm d}$ to be randomly sampled up until the Toomre inequality (which is set as the upper boundary, $Q {>} 1$).  The Toomre criterion is defined as: 

%\begin{equation}
 %   Q_{\rm gas} {=} \frac{c_{\rm s} \kappa}{\pi G \Sigma},
%\end{equation}

%where $c_{\rm s}$ is the speed of sound (a measure of thermal pressure); $\kappa$ is the epicyclic frequency ($\kappa{=}\Omega$ in a Keplerian disk); $G$ is Newton's gravitational constant; and $\Sigma$ is the surface density of the disk.  We choose the Toomre instability criterion as the upper bounds of this parameter as this condition must be satisfied for the disk to be stable against collapse, any more massive causes disk fragmentation - leading to gravitational instability - which is beyond the scope of this paper as our focus is solely on giant planet formation through pebble accretion. 

\subsection{Stellar sample}
\label{sec:sample}

\begin{figure}
    \includegraphics[width=\linewidth]{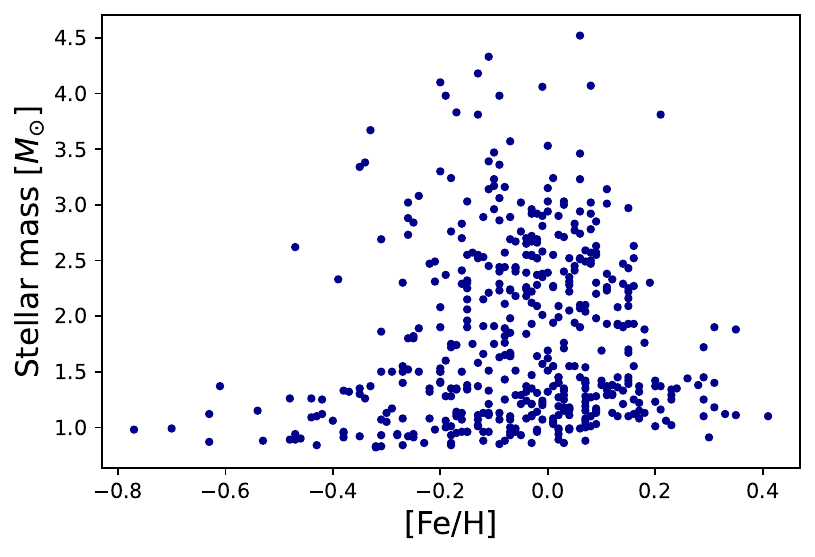}
    \caption{Stellar metallicity versus mass for the 482 stars of the homogenised combined RV sample \citep{Wolthoff2022PreciseSurveys}.}
    \label{fig:w22_stars}
\end{figure}

For our analysis, we use the homogenised stellar sample in \cite{Wolthoff2022PreciseSurveys}, which is comprised of 482 stars (see Figure \ref{fig:w22_stars}).  This combines observations from three RV surveys: the Lick giant star survey \citep{Reffert2006PRECISECOMPANION}; the EXo-Planets aRound Evoled StarS (EXPRESS) survey \citep{Jones2011StudyProperties}; and the Pan-Pacific Planet Search (PPPS) \citep{Wittenmyer2011ThePeriods}.  This sample accounts for partial target overlaps of the three surveys (EXPRESS  and PPPS surveys have 37 stars in common; Lick and EXPRESS share 12 stars; and Lick and PPPS overlap with one star) and contains only single star systems.  %As in \cite{Wolthoff2022PreciseSurveys}, we also disregard stars of masses $\geq 5 \Ms$, high luminosities ($L_{\star} \geq 10^{3} \rm L_{\odot}$), are most likely supergiants ($\log (g) \leq 1.5$), or have only a few observations (\#RVs $ < 10$).  Overall, this leaves 482 stars for our occurrence rate analysis (see Figure \ref{fig:w22_stars}).

The stellar parameters (mass, radius, age, surface gravity, effective temperature, and luminosity) for the extended sample were homogeneously derived by \cite{Wolthoff2022PreciseSurveys} using the method of  \cite{Stock2018PreciseSearch}, which uses Bayesian inference to compare  spectroscopic, photometric, and astrometric observables to the PARSEC evolutionary tracks \citep{Bressan2012Parsec:Code}, incorporating evolutionary timescales and the initial mass function as priors.  For each stellar parameter, two probability distribution functions are generated: one assuming the star is ascending the red-giant branch (RGB), and one assuming it has reached the horizontal branch (HB), each with an associated probability.  \cite{Wolthoff2022PreciseSurveys} adopt the stellar parameters corresponding to the most probable evolutionary stage.

Concerns have been raised in the literature about the systematic overestimation of stellar masses \citep{Lloyd2011RetiredLunch, Lloyd2013TheHosts, Schlaufman2013EvidenceStars}.  However, several studies found no evidence for a significant systematic discrepancy \citep{Johnson2013RetiredSubgiants, Ghezzi2015BEYONDTRACKS, North2017TheHosts, Ghezzi2018RetiredMass}.  The method of \cite{Stock2018PreciseSearch} is robust in this regard, as it explicitly accounts for evolutionary timescales, the initial mass function, and the two distinct evolutionary stages separately.

%has demonstrated their robustness in this regard as they specifically  account for these evolutionary mass scales, initial mass function, and fits the red-giant branch (RBG) and the horizontal branch (HB) separately.

%The Bayesian inference method of \cite{Stock2018PreciseSearch}  generates two probability distribution functions for each of the stellar parameters listed above: (i) the first assuming the star is either ascending the RGB or (ii) the star has reached the HB, combined with an estimated probability for each of these two evolutionary stages.  \cite{Wolthoff2022PreciseSurveys} used the stellar parameters for the most likely evolutionary stage.
%In the analysis by \cite{Wolthoff2022PreciseSurveys}, they use the mode values of the distributions to estimate a more probable evolutionary stage.  

%\textbf{Comment whether there may have been (or not!) any $M_{\star}$ loss since PMS. (???)} 

%- trends, findings from last paper
%- disk mass assumption 
\subsection{Accretion rate}
\label{sec:acc}
%- updated accretion plot, findings from observations, findings from the last paper (theoretical expectation was to be quadratic but this is not what is observed, discrepancy with time 0 etc. ).  Synthesis will proceed for M$17$, W$20$, some slop in between. 

Mass accretion rate is one of the most important factors in any kind of planet formation simulations.  It governs the rate of dust and gas fall onto the central star (and hence disk lifetime) as well as how efficiently pebble accretion can proceed.  It is especially important in the case of giant planet formation as the accretion rate must be high enough in order for planets to grow massive before the disk material is dissipated through photoevaporation processes \citep{Liu2019Super-EarthMasses, Johnston2024FormationStars}. 

\begin{figure}
\includegraphics[width=\linewidth]%{Accr_Herbigs_TTauris.pdf}
%{accrate_age_comp.pdf}
%{accrate_age_fits.pdf}
{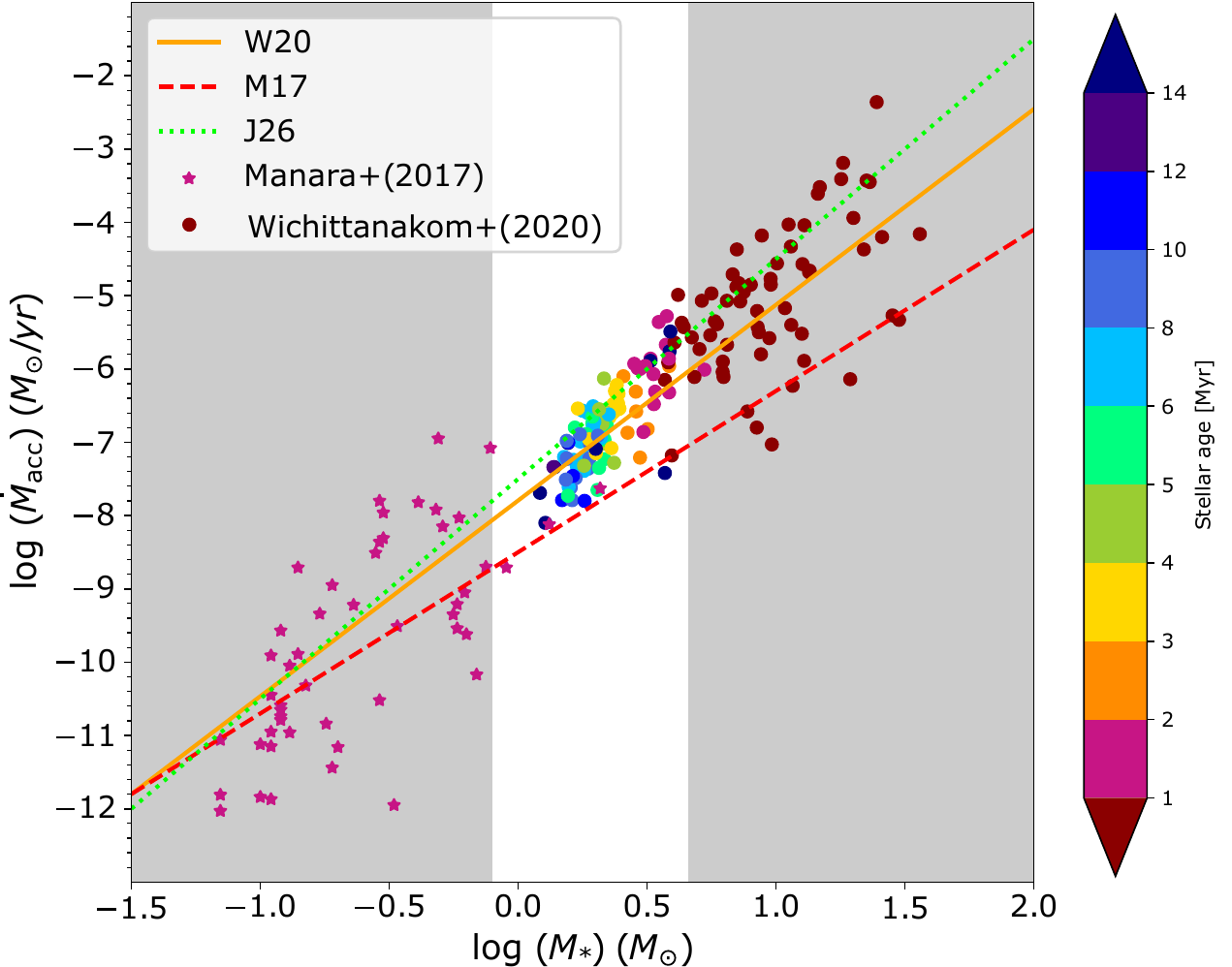}
\caption{Mass accretion rates from observational studies \citep{Wichittanakom2020TheStars, Manara2018WhyPopulation}, adapted from the Appendix of \citep{Iglesias2023X-ShooterEvolution}.  Linear fits are shown for reference; dashed red line for M$17$, dotted lime for J26, and solid orange line for W$20$.  The stars are colour coded by stellar age.  There are no individual ages for stars in the sample by \citep{Manara2017X-ShooterStars}, so we adopt the age of the Chamaeleon I star-forming region to be 1.5 Myr from \citep{Galli2021ChamaeleonData}.  The white highlighted region is the stellar mass range that we are interested in exploring.}%  There are no individual ages for \citep{Manara2023DemographicsFormation} sample so the age of the star-forming regions were adopted instead.}
\label{fig:accage}
\end{figure}

Figure \ref{fig:accage} is adapted from the Appendix of \cite{Iglesias2023X-ShooterEvolution} and shows the mass accretion rates for the Herbig Ae/Be stars in \cite{Wichittanakom2020TheStars, Iglesias2023X-ShooterEvolution}, along with the sample of low-mass stars in the Chamaeleon I star-forming region studied in \cite{Manara2017X-ShooterStars}.  The highlighted region shows our stellar mass range of interest.  The best linear fits for each sample is shown, derived over their respective full mass ranges: M17 refers to the fit derived from \cite{Manara2017X-ShooterStars}; W20 to that from \cite{Wichittanakom2020TheStars}; and J26 to our own estimate as described below.  When we are conducting our planet population synthesis, we use the recorded stellar masses from our sample in Figure \ref{fig:w22_stars} \citep{Wolthoff2022PreciseSurveys}, and select the corresponding accretion rate $\dot{M}_{\rm acc}$ from the linear fit of interest in Figure \ref{fig:accage} and use that as the initial accretion rate $\dot{M}_{\rm g0}$ at $t {=} 0$ Myr in Equation \ref{eq:mdotg} for our model.  %We use these best single-line fits to derive the respective accretion rates for the stars in our sample (see Figure \ref{fig:w22_stars}) to conduct planet population synthesis.  }  %where the average age of the systems are $1{-}2$ Myr \citep{Galli2021ChamaeleonData}.

Furthermore, we consider the ages of the stars in Figure \ref{fig:accage}, where the colour of the data is associated with the stellar age in Myr.  There are no individual ages for the low-mass stars in the \cite{Manara2017X-ShooterStars} sample so we instead adopt the age of the Chamaeleon I star-forming region to be 1.5 Myr \citep{Galli2021ChamaeleonData}.  The ages for the Herbig Ae/Be stars in \cite{Wichittanakom2020TheStars} are either derived from the same study or from \cite{Vioque2018GaiaStars}. The highlighted region again denotes our stellar mass range of interest for this study. The majority of the \cite{Manara2017X-ShooterStars} low-mass stars, alongside the very young and massive Herbig Ae/Be stars in the \cite{Wichittanakom2020TheStars} sample, are beyond our stellar mass range of interest. 

The stars with masses that do lie in our range of interest are mainly Herbig Ae/Be stars in the \cite{Iglesias2023X-ShooterEvolution} and a part of the \cite{Wichittanakom2020TheStars} sample.  Such stars vary in age from $1{-}20$ Myr, with only a few of the most massive stars having ages reported below 1 Myr.  This diversity in the age of these stars provides an interesting challenge for us given that accretion rates decay over time, typically beginning to drop substantially between $1{-}2$ Myr for $1{-}2.4 \Ms$ stars \citep{Johnston2024FormationStars}.  Using observed accretion rates for older stars as initial conditions at $t=0.1$ Myr would therefore underestimate the true initial accretion rate.  To account for this, we derive a new accretion rate relation (J26) by fitting a linear relation to the accretion rates of intermediate-mass stars in the \cite{Wichittanakom2020TheStars} and \cite{Iglesias2023X-ShooterEvolution} samples, accounting for the wide range of ages present in these samples. This fit provides a steeper relation that better represents the initial conditions at $t=0.1$ Myr.

\subsection{Disk mass}
\label{sec:diskmass}

As in \cite{Pascucci2016ARELATION}, we assume the disk mass to be super-linearly correlated with the mass of the host star.  This is determined by dust continuum emission measurements with ALMA \citep{Pascucci2016ARELATION, Manara2016EvidenceDisks}.  In particular, \cite{Manara2016EvidenceDisks} combined results of surveys from VLT/X-shooter and ALMA to find a correlation between the mass accretion rate and the disk dust mass, confirming that mass accretion rates are related to the properties of the outer disk.  In our model, the disk evolves viscously in tandem with the mass accretion rate ($\dot{M}_{\rm g}$) onto the central star and is correlated with the mass of the disk ($M_{\rm d}$) \citep{Hartmann1998ACCRETIONNURIA}.  Hence, the ratio between $\dot{M}_{\rm g}$ and $M_{\rm d}$ can be expressed in terms of the viscous timescale at the outer edge of the disk \citep{Liu2019Super-EarthMasses}: 

\begin{equation}
\label{eq:tvis}
    \tau_{\rm vis}{=} \frac{1}{3(2 + s)^{2}} \frac{R_{\rm d0}^{2}}{\nu_{0}}.
\end{equation}

$R_{\rm d0}$ and $\nu_{0}$ are the initial characteristic size of the gas disk and the viscosity at that radius, and $s$ is the gas surface density gradient.  Notably, the viscosity is dependent on sound speed $(c_{\rm s})$, scale height of the gas disk $(H_{\rm g})$, and Keplerian angular frequency $(\Omegak)$, as defined above.  These terms have $M_{\star}$- and $L_{\star}$-dependence, meaning that the viscous timescale evolves differently around host stars of different masses.  %These are intrinsically dependent on the mass of the host starand irradiated temperature of the disk (and hence the luminosity of the host star, $L_{\star}$).  This results in the viscous timescale $(\tau_{\rm vis})$ - and the viscous dominated accretion rate regime - evolving differently around different masses of host stars.  } 

%\subsection{Disk size}
Initial characteristic disk size - $R_{\rm d0}$ - is the corresponding length scale for viscous disk evolution \citep{Liu2019Super-EarthMasses}.  %This is the assumption used to achieve the first set of giant planet occurrence rate calculations.
\cite{Bate2018OnDiscs} performed sophisticated radiation hydrodynamical simulations of star formation in clusters. Their results indicated that the sizes of the early protostellar disks are poorly (or at most weakly) dependent on $M_{\star}$. This is consistent with observational studies, which have also found no clear correlation between disk size and stellar mass (\textit{e.g.} \cite{Andrews2013TheHosts, Hendler2020TheStars}). However, we note that recent findings by \cite{Stapper2026AnKpc} suggest an increase of dust disk mass with stellar mass up to $1{-}2$ M$_{\odot}$; and then a decrease with stellar mass, likely related to an increase in UV radiation, multiplicity, and/or radial drift. 
%However, \cite{Stapper2022TheALMA} found that disks around Herbig stars were generally both larger and more massive than disks around T Tauri stars, but with the caveat that the largest Herbig disk is of similar size to the largest T Tauri disk. They also speculate that these massive and large disks originate from initial disk mass and subsequent disk evolution. 
Nonetheless, we assume that $R_{\rm d0}$ has no correlation with $M_{\star}$.  Instead, we allow $R_{\rm d0}$ to be sampled uniformly between $20{-}200$ au.

We note that, since $R_{\textrm{d0}}$ is independent of stellar mass in our model, prescribing $\dot{M}_{\textrm{g0}}$ as a function of $M_{\star}$ is equivalent to setting the initial disk mass $M_{\textrm{d}}$, as the two are related through the viscous timescale (Equation \ref{eq:tvis}).  The J26 model therefore implies an initial disk mass relation $M_{\textrm{d}}\propto M_{\star}^{3}$, which is steeper than the relations derived from millimetre continuum surveys of Class I/II disks: $M_{\textrm{d}} \propto M_{\star}^{1}$ \citep{Williams2011ProtoplanetaryEvolution} and $M_{\textrm{d}}\propto M_{\star}^{1.3{-}1.9}$ \citep{Pascucci2016ARELATION}.  However, these observational relations are inferred at typical disk ages of $1{-}3$ Myr, corresponding to Class I/II evolutionary stages.  Our model is initialised at $t=0.1$ Myr, representing Class 0 or early Class I conditions, at which point disks around more massive stars have had less time to drain through viscous accretion and photoevaporation.  Since both processes are more efficient around higher mass stars \citep{Ribas2015ProtoplanetaryPopulations, Komaki2020RadiationRate}, the disk mass and stellar mass relation is expected to be steeper at these earlier epochs and to flatten progressively as the disk evolves. A steep initial scaling is therefore physically consistent with our Class 0 initial conditions and with the observed Class I/II relation representing the outcome of subsequent evolution.  Observational constraints on disk masses at Class 0 stages remain scarce, making this assumption difficult to test directly, but represents a physically motivated initial condition for our model.

We further note that these observational relations should be interpreted with care when compared to our model.  The \cite{Williams2011ProtoplanetaryEvolution} relation, $M_{\textrm{d}} \propto M_{\star}^{1}$, was derived using massive stars and is therefore more appropriate for the intermediate-mass sample considered here than the \cite{Pascucci2016ARELATION} relation, which is only statistically significant up to 2 M$_{\odot}$ and falls short of the stellar masses considered in this work. Furthermore, \cite{Iglesias2023X-ShooterEvolution} demonstrate that Herbig Ae/Be stars are not representative of intermediate-mass star discs as a whole due to their advanced ages (average ages $\sim 6$ Myr).  There is currently no strong observational evidence for the disc mass - stellar mass relation above 2 M$_{\odot}$.

\subsection{Birth location and birth embryo mass}

The birth location of the embryo is randomly sampled between $0.1{-}100$ au.  This has important consequences for the initial birth mass of the embryo as we adopt our streaming instability condition, as defined in Equation \ref{eq:embryo_mass}.  

\begin{figure}
    \includegraphics[width=\linewidth]{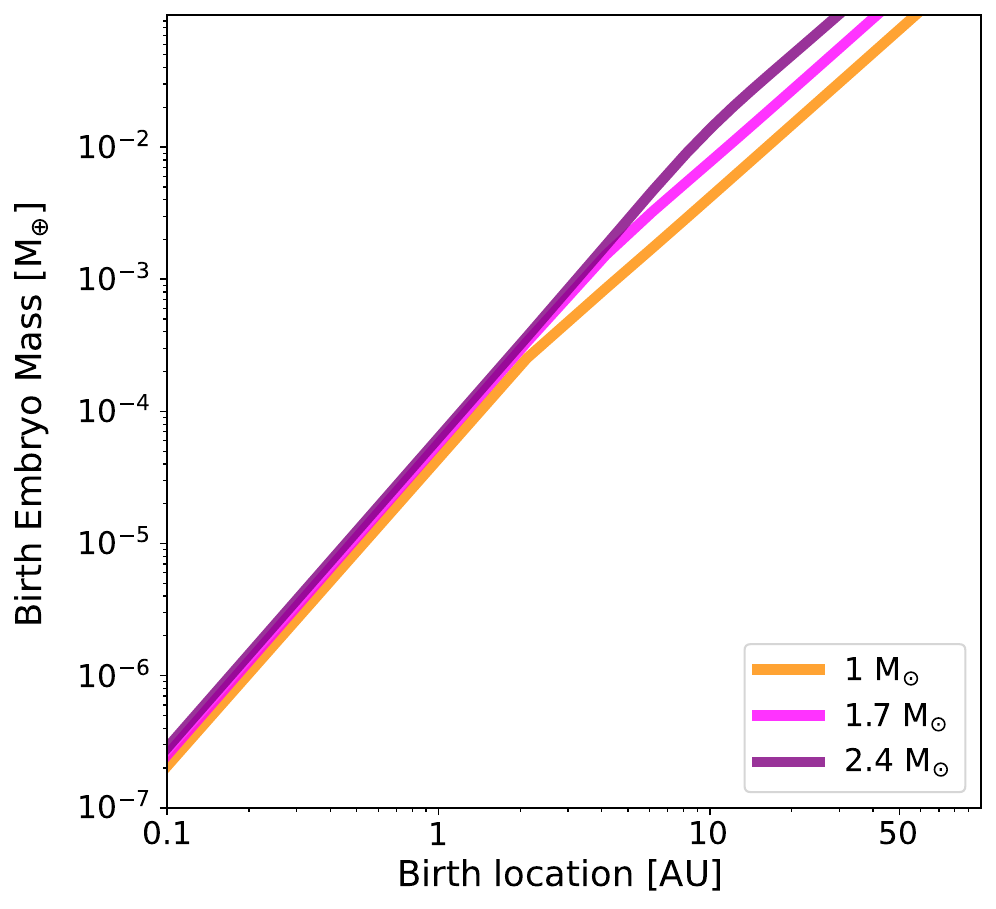}
    \caption{Birth location against birth embryo mass for time 0 for key stellar masses 1 (orange), 1.7 (pink), and 2.4 $\Ms$ (purple).  Within the first few au of the host stars, the birth embryo masses are essentially identical. The breaks in each line correspond to the transition radius $r_{\textrm{tran}}$ between the viscously heated and stellar irradiated disk regions (Equation \ref{eq:rtrans}).  The J26 accretion rate model was assumed for this plot.  }
    \label{fig:varemb}
\end{figure}

Figure \ref{fig:varemb} demonstrates the initial birth mass of embryos at 0.1 Myr in disks around stars of masses 1, 1.7, and 2.4 $\Ms$.  Within the first few au, the birth masses are essentially identical regardless of stellar mass.  These stellar masses are illustrative but correspond to the rise, peak, and fall of the observed occurrence rate.  In this mass range, the stars remain wholly convective (\textit{i.e.} dominated by X-ray photoevaporation) for the times taken for giant planet formation to conclude in our models.  At birth locations beyond $3$ au, the birth masses begin to diverge, with the largest birth masses corresponding to the most massive stars.  This in turn impacts the rate of pebble accretion (Equations \ref{eq:peb_acc} \& \ref{eq:peb_acc_eff}) and subsequently dictates when/if runaway gas accretion can proceed.  

\subsection{Birth time}
\label{sec:birth_time}

We set the birth time for all embryos to be 0.1 Myr.  While \cite{Johnston2024FormationStars}found that giant planet formation can proceed up to 3 Myr under the most favourable conditions, we fix the embryo formation time to the earliest epoch in order to circumnavigate the limitation of our model: we employ only an X-ray photoevaporation prescription, despite our evolving PMS (Section \ref{sec:pms_ev}).  We do not model the transition in photoevaporation mechanism from X-ray to FUV \citep{Kunitomo2021PhotoevaporativeStars},   which occurs when a convective star becomes radiative.  For stars less massive than 2.5 M$_{\odot}$, - the range corresponding to the rise, peak, and fall of the giant planet occurrence rate as a function of stellar mass - disks are still primarily dispersed by X-ray photoevaporation \citep{Kunitomo2021PhotoevaporativeStars}.  Additionally, the photospheres of stars less massive than $1.6 \Ms$ never become hot enough to emit FUV \citep{Kunitomo2021PhotoevaporativeStars}.  By fixing $t_{0}=0.1$ Myr, our embryos are formed when X-rays remain the dominant mechanism across most of our stellar mass range.

%Furthermore, for stars less massive than $2.5 \Ms$, \textbf{their discs} are still mainly dispersed by X-ray photoevaporation \citep{Kunitomo2021PhotoevaporativeStars}, and this is where the main stellar mass range of interest corresponds to the rise, peak, and fall of the giant planet occurrence rate as a function of stellar mass. 

\begin{table}
    \centering
    \begin{tabular}{lclclclclclc|}
        \hline
        \hline
        Parameter   &   Description  &   \\ 
            \hline
        disk model   &  viscously heated + stellar irradiation \\ 
        $\rm M_{\star}$ [$\Ms$]  & Figure \ref{fig:w22_stars} sample $(0.82{-}4.52)$  \\ 
        $\rm Z_{\star}$ [Fe/H] & Figure \ref{fig:w22_stars} sample $(-0.77{-}0.41)$ \\
         $\dot M_{\rm g0}$  [$ \rm \Ms \ yr^{-1}$] & Derived from Figure \ref{fig:accage}.  \\
         & J26: $y{=}3 x - 7.5$ \\
         & W20: $y{=}2.67 x - 7.8$; \\
         & M17: $y{=}2.2 x - 8.5$ \\
         $ R_{\rm d0}$ [AU] &  $\rm{ U}(20,200)$ \\
          $ r_0$ [AU] &  $\rm{\log U}(0.1,100)$ \\
           $ t_0$ [Myr] & $0.1$ \\ 
            $R_{\rm peb}$ [mm] &  $1$ \\
            $ M_{\rm p0} [\Me] $ &  \eq{embryo_mass}, Figure \ref{fig:varemb} \\
      \hline
        \hline
    \end{tabular}
\caption{Adopted parameter distributions for the population synthesis study.}
    \label{tab:pps}
\end{table}

\section{Giant planet occurrence rate as a function of stellar mass and metallicity}
\label{sec:results}

%\citep{Reffert2015PreciseMetallicity} obtained the precise RVs for a sample of 373 bright ($V \leq 6$ mag) G- and K- type giants at the \textit{Lick Observatory} for about 12 years.  The selection process is detailed in \citep{Frink2001AMission}. 

One of the most striking results from \cite{Reffert2015PreciseMetallicity} was the steep decline in the giant planet occurrence rate for stellar masses greater than $2.5{-}3\ \Ms$.  Of the 113 stars examined, no confirmed giant planets were found around stars more massive than $2.7 \Ms$.  Thus, their planet occurrence rate for stars in the range of $2.7{-}5\ \Ms$ was found to be $0.0^{+1.6}_{-0.0}\%$ \citep{Reffert2015PreciseMetallicity}.  This complements earlier works which found an increase in the occurrence rates of giant planets with $M_{\star}$, but only exploring up to 2 M$_{\odot}$ \citep{Johnson2010GIANTPLANE, Bowler2010RetiredStars}.  

\cite{Wolthoff2022PreciseSurveys} repeated the same process by combining data from the Lick, EXPRESS, and PPPS giant star RV surveys.  This sample consisted of 482 stars hosting 37 giant planets in 32 systems, with planetary masses in the range of $0.8{-}24.4\ \rm M_{\rm Jup}$.  The majority of these planets were concentrated at higher metallicities and stellar masses between $1{-}2 \Ms$, with no planet-hosting stars more massive than $3 \Ms$ \citep{Wolthoff2022PreciseSurveys}.  They also found a distinct peak in the giant planet occurrence rate above $1.5\Ms$ with an exponential drop for stars more massive than 1.8 M$_{\odot}$, consistent with \cite{Reffert2015PreciseMetallicity, Jones2015Giant67851c}.  During their Bayesian fitting, they found the maximum planet occurrence rate is reached at $M_{\star}{=}1.7 \Ms$.  

 Taking selection biases and observational sensitivity into account, \cite{ Wolthoff2022PreciseSurveys} then fit the giant planet occurrence rate as a function of stellar mass and metallicity with a Gaussian and exponential fit, respectively.  The Gaussian plus exponential distribution has the following from with the model parameters $X=(C, \mu, \sigma, \beta)$: 

\begin{equation}
\label{eq:fit}
    f(M_{\star}, [\rm Fe/H]) = C \exp \left( - \frac{1}{2} \left[ \frac{M_{\star} - \mu}{\sigma} \right]^{2} \right) 10^{\beta [\rm Fe/H]} ,
\end{equation}

where the maximum likelihood values from \cite{Wolthoff2022PreciseSurveys} are: $C{=}0.128 \pm 0.027$, $\mu{=}1.68 \pm 0.59 \Ms$, $\sigma{=}0.70 \pm 1.02 \Ms$, and $\beta{=}1.38 \pm 0.36$.

%In this section, we present the results from our planet population synthesis.  We compare how our \textbf{accretion rates} from Figure \ref{fig:accage} correspond to the observed giant planet occurrence rate reported in \cite{Wolthoff2022PreciseSurveys}.  We also investigate any trends in the initial conditions of the embryos that form giant planets. 

\subsection{Dependence of the giant planet occurrence rate on the accretion rate }

\begin{figure*}
    \includegraphics[width=\linewidth]%{heatmap_comb.pdf}
    %{pps_heatmap_comb_tot.pdf}
    {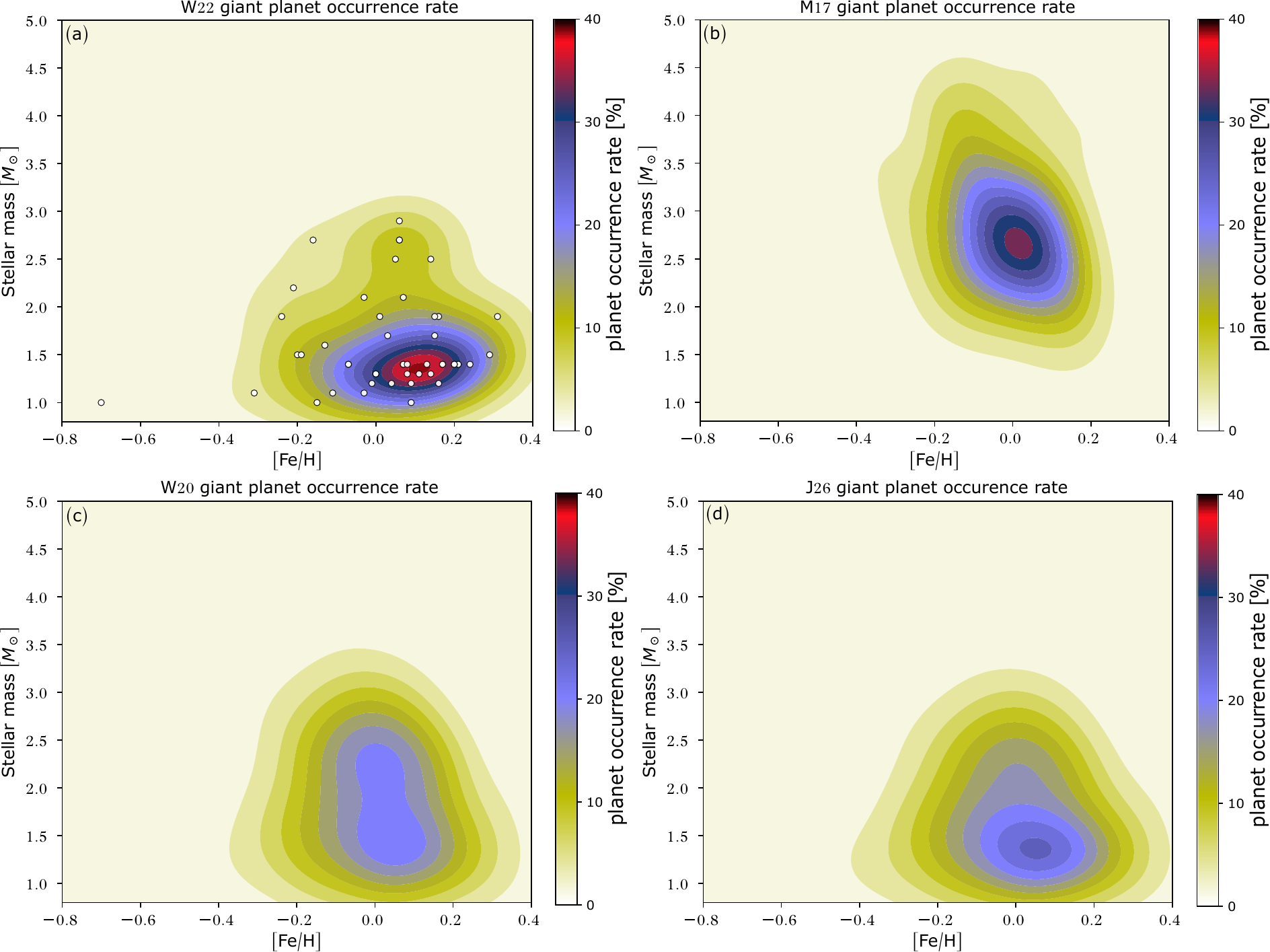}
    \caption{Planet occurrence rate (colour coded) as a function of metallicity and stellar mass for the stars in \citep{Wolthoff2022PreciseSurveys} (a) and in our synthetic models: M17 (b), W20 (c), and J26 (d).   Data has been adapted from \citep{Wolthoff2022PreciseSurveys} and heavily smoothed in (a) so that general trend can be examined with the exact observations shown as white circles.  There is a clear maximum in the observed planet occurrence rate (a) for metallicities of about 0.2 and masses of about $2 \Ms$.  For our models (b, c, d), the peak of planet formation shifts depending on the initial conditions and is more concentrated as the data does not need to be smoothed.}
    \label{fig:heat}
\end{figure*}

We will now examine the results of giant planet occurrence rate in relation to the stellar mass and metallicity in detail.  We do this by counting the number of giants produced from our population synthesis per stellar mass/metallicity combination corresponding to the values from our RV sample in Figure \ref{fig:w22_stars} from \cite{Wolthoff2022PreciseSurveys}.  We do this for each of our stellar mass-accretion rate relations outlined in Figure \ref{fig:accage} (M17, W20, and J26), such that we can examine the importance of the mass accretion rate in modelling the observed giant planet rate as a function of stellar mass.  

In Figure \ref{fig:heat}, planet occurrence rates are colour-coded according to the bar on the right hand side of the plots.  %The planet occurrence rate ranges from $0\%$ (white) up to $40\%$ (black).  
The blank areas of the diagrams correspond to metallicity/mass combinations with no existing measurements in the observational sample (Figure \ref{fig:w22_stars}).  

Figure \ref{fig:heat} shows the planet occurrence heat maps from our models alongside the observed giant planet occurrence rate for the \cite{Wolthoff2022PreciseSurveys} sample.   It shows a clear peak in the giant planet rate for high metallicities of $\sim 0.2$ dex and stellar masses around $\sim 1.7 \Ms$.

%\subsubsection{M17}
%\label{sec:map_m17}

Figure \ref{fig:heat}b shows the heat map for the planet occurrence rate for the M17 accretion rates shown in Figure \ref{fig:accage}.  The occurrence rate of giant planets very clearly increases with increasing stellar mass and metallicity.  Occurrence rates $\geq 20\%$ are only possible when the stellar mass is $\geq 3 \Ms$,  which is clearly at odds with the observational peak of $M_{\star} \sim 1.7 \Ms$ calculated in \cite{Wolthoff2022PreciseSurveys}.  %We will consider the parameters in our model that are contributing to this disconnect below.}
%but there is a clear factor behind the broadening and shifting of the occurrence rates - the overall low accretion rate fit of M17.  

For our W20 model in Figure \ref{fig:heat}c, we see that the mass/metallicity combinations required for high occurrence rates are very different than in Figure \ref{fig:heat}b.  The occurrence rates of around $20\%$ (blue) are now found at stellar masses $\leq 2.5 \Ms$ and metallicities between  $-0.2{-}0.2$.  There is some evidence of giant planet occurrence rates of $20\%$ (blue) up to $3 \Ms$ for only high metallicities.  The peak of giant planet occurrence distribution (more than $30\%$, dark blue) now occurs around stellar mass of $1.7 \Ms$ and metallicities $0{-}0.2$.  %Our modelled planet rates are  similar to that of the observational occurrence rates in \cite{Reffert2015PreciseMetallicity} (Figure \ref{fig:heat}a). 
W20 produces lower occurrence rates than W22 in some bins, which we attribute to the smaller sample size of W22 leading to elevated rates in individual bins due to small-number statistics.  %The W20 model has a much finer parameter space due to the nature of the model described in Section \ref{sec:method}.  
However, both W20 and W22 are agreement that the highest rates of giant planets are found in systems with M$_{\star}=1.5-2$ M$_{\odot}$ and metallicity $\sim 0.2$ dex.   

Finally, we consider our J26 model of giant planet rate in Figure \ref{fig:heat}d.  Overall, it is broadly similar to both the W20 model and the W22 planet rate, albeit more concentrated in the region of high metallicities and stellar masses $\sim 1.6{-}1.8 \Ms$.    

As discussed in Section \ref{sec:acc}, the M17 fit is the best fit for the low-mass T Tauri stars in the \citep{Manara2017X-ShooterStars} sample.  Given that the majority of our stellar mass sample from \cite{Wolthoff2022PreciseSurveys} are more massive than $ 1\Ms$, this results in the accretion rates for our model systems to be lower than one might observe in low-mass stars.  Furthermore, when we consider the ages of the stars in the mass accretion rate observations in Figure \ref{fig:accage}, it is clear that this accretion rate fit M17 is not representative of accretion rates for stars with advance ages ($\geq 10$ Myr) for the upper end of stellar masses in our sample.  The consequence of these low observed initial accretion rates around more massive stars is the prolonging of the protoplanetary disk lifetime.  This results in the model giant planet occurrence rate being skewed to higher stellar masses because there is a weaker limiting factor of the disk lifetime in such systems.     

Concurrently, there are extremely low giant planet occurrence rates when the stellar masses are less than $2 \Ms$.  Again, a consequence of the very low accretion rates from the less massive stars in our sample.  While the disks around stars less massive than $2 \Ms$ likely survive for a long time, the gas accretion is so low that pebble accretion (Equation \ref{eq:peb_acc}) becomes incredibly inefficient.  Many of the planets in such systems would not be able to accrete enough solid material to undergo runaway gas accretion.  This is further exacerbated by our assumption within the planet formation model that the disk mass is correlated with the mass accretion rate (described in Section \ref{sec:diskmass}).  Hence we can conclude that this particular model setup is not a good match for the observed giant planet occurrence rate and do not include in further analysis.

%When we sample accretion rates along the W20 fit \ref{fig:accrate} (orange), we get an outcome that is a closer fit to the observational results.  The giant planet occurrence rate of $\sim 30\%$ is once again concentrated in metallicities of $0.1{-}0.2$ and stellar masses of $\sim 1.9 \Ms$.  

%\subsubsection{W20}
%\label{sec:map_w20}

%Now, we explore the planet occurrence rate when we conduct our population synthesis by sampling accretion rates from W20 in Figure \ref{fig:accrate}.  In Figure \ref{fig:heat}c, we see that the mass/metallicity combinations required for high occurrence rates are very different than in Figure \ref{fig:heat}b.  The occurrence rates of around $20\%$ (blue) are now found at stellar masses $\leq 2.5 \Ms$ and metallicities between  $-0.2{-}0.2$.  There is some evidence of giant planet occurrence rates of $20\%$ (blue) up to $3 \Ms$ for only high metallicities.  The peak of giant planet occurrence distribution (more than $30\%$, dark blue) now occurs around stellar mass of $1.8 \Ms$ and metallicities $0{-}0.2$.  Our modelled planet rates are very similar to that of the observational occurrence rates in \cite{Reffert2015PreciseMetallicity}.  

The W20 and J26 models have initial accretion rates are much higher than the M17 fit overall.  This impacts where the peak of the giant planet occurrence rate occurs: (i) for less massive stars, the higher accretion rates improves the rate of pebble accretion so they can surpass pebble isolation mass to undergo runaway gas accretion at earlier times and (ii) for more massive stars,  the very high accretion rates aids in fast disk evolution which suppresses the timeframe for giant planet assembly. 
 %We explore these two models further in latter sections.

To conclude, the final planet occurrence rate is very sensitive to the accretion relation, and thus stellar-mass dependent accretion rates are vital to giant planet formation. There is a balance to be struck between efficient pebble accretion and dissipating the disk too quickly.  We explore this balance further below.

\subsection{Giant planet dependence on M$_{\star}$ and Fe/H}
\label{sec:func}

%\begin{figure}
%    \includegraphics[width=\linewidth]{occ_rate_hist_z.pdf}
%    \caption{Histogram of our mean model giant planet occurrence rates as a function of stellar metallicity. Our mean model data is coloured green (J23); red (\citep{Manara2017X-ShooterStars} estimation); or orange (\citep{Wichittanakom2020TheStars}).  The filled red histogram shows secure giant planets in the \textit{Lick} sample \citep{Reffert2015PreciseMetallicity}. The black line denotes the best fit to mass dependence for $1.9 \Ms$ in the observational data \citep{Reffert2015PreciseMetallicity}.  The black dots correspond to the same model but the true stellar mass distribution within each bin has been taken into account \citep{Reffert2015PreciseMetallicity}.  }
%    \label{fig:hist_z}
%\end{figure}

\begin{figure*}
    \includegraphics[width=\linewidth]
    {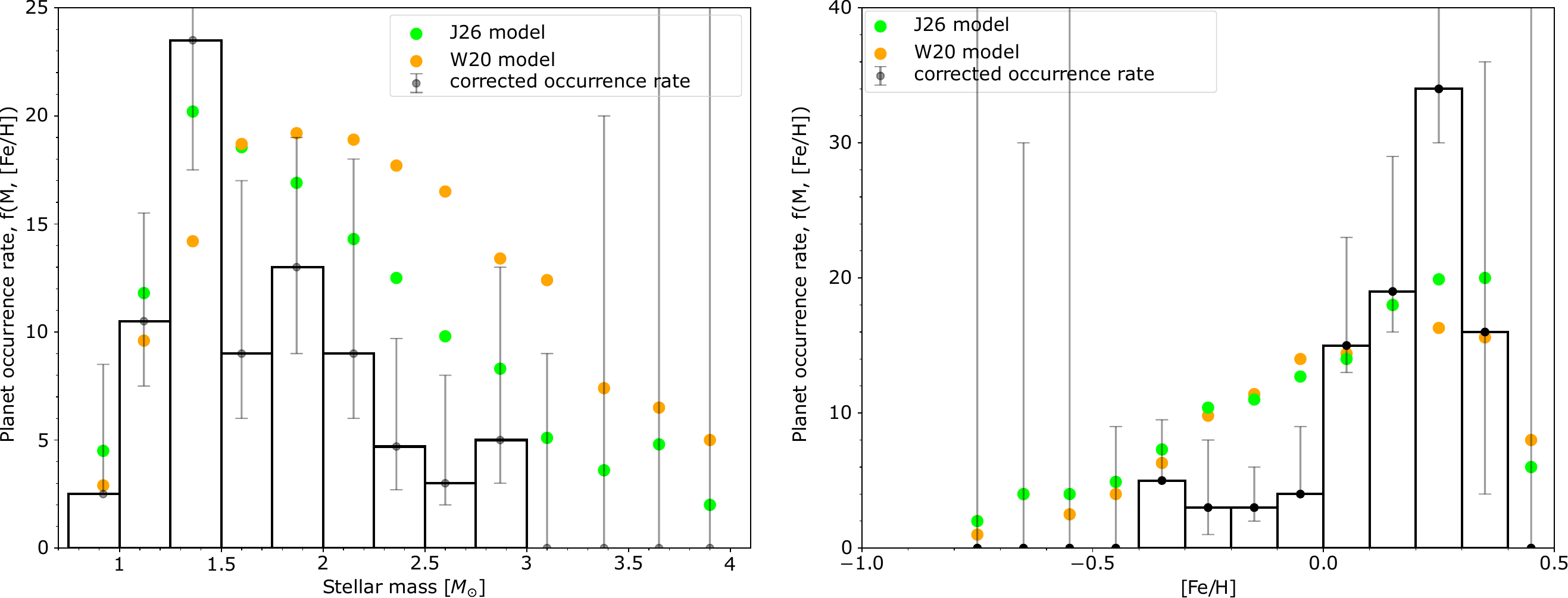}
    %{occ_rate_hist_jul24.pdf}%{occ_rate_hist_comb_tot_new.pdf}
    
    \caption{Histogram of our mean model giant planet occurrence rates as a function of stellar mass (left) and metallicity (right). Our best mean models are coloured in orange (W20) and lime (J26), respectively.  
    %The red dotted histogram shows the distribution of planets \citep{Wolthoff2022PreciseSurveys}, while 
    The solid black histograms shows the completeness-corrected observed giant planet occurrence rate in each bin \citep{Wolthoff2022PreciseSurveys}.  
    %The blue dotted line is derived from Equation \ref{eq:fit} and is the fitted occurrence rate as a function of stellar mass (left) with the metallicity set to the sample mean of $\rm [Fe/H]{=}0.05$ and as a function of metallicity (right) with the mean stellar mass set to $M_{\star}{=}1.81 \Ms$.  
    %The histogram is for visualisation, the fit is independent of the binning of the data. 
    }  
    \label{fig:hist}
\end{figure*}

In Figure \ref{fig:hist}, we show the mean model giant planet occurrence rate as a function of stellar mass (left)  and metallicity (right).   %We follow the procedure in \cite{Wolthoff2022PreciseSurveys} to derive the binned histograms for the observed planet rate (red, dotted) and the corrected occurrence rate (black, solid) for their observational sample in Figure \ref{fig:w22_stars}.  The blue dotted line is the fitted occurrence rate as a function of stellar mass and metallicity, also from \cite{Wolthoff2022PreciseSurveys}, and defined in Equation \ref{eq:fit}.  
Our mean-averaged results for for the W20 and J26 models, derived using the same binning process, are shown in orange and lime respectively.

When binned as a function of stellar mass in Figure \ref{fig:hist} (left), we can see that our W20 model follows a broadly similar shape to the observed planet rate.  The peak of our model planet distribution is found when the mean stellar mass bin is $M_{\star}{=}1.87 \Ms$.  Overall, the giant planet occurrence rate reaches $\eta_{\textrm{J}} \sim 19$ \% for stars in the range of $1.6{-}2.15$ M$_{\odot}$. 
There is a slow decline in the planet rate for stellar masses $M_{\star} > 2.15 \Ms$ which drops off more steeply when $M_{\star} > 2.5 \Ms$.  Although consistent with the observational findings of a distinct peak in the occurrence rate for stars more massive than  $1.5 \Ms$ \citep{Reffert2015PreciseMetallicity, Jones2015Giant67851c, Wolthoff2022PreciseSurveys}, our drop-off in stars more massive than $M_{\star} > 2.5 \Ms$ is not quite as steep as the observed results.  Our planet occurrence rate remains above $5\%$ for even the most massive stars in our sample.  

The decrease in the giant planet occurrence rate for stars more massive than $\sim 2$ M$_{\odot
}$ is consistent with the shorter disk lifetimes expected around these stars.  Observational studies suggest that disks around intermediate-mass stars are dissipated within a few Myr (\textit{e.g.} \cite{Iglesias2023X-ShooterEvolution}), which limits the time available for giant planet formation.  This disk-lifetime mechanism has been invoked by \cite{Burkert2007THEM, Currie2009OnStars}, and \cite{Alibert2011ExtrasolarMasses} to explain the paucity of close-in giant planets around intermediate-mass stars. 
%\cite{Ribas2015ProtoplanetaryPopulations} draw a line at 3 M$_{\odot}$, and almost all of our sources fall under 3 M$_{\odot}$ - all considered 'low-mass' in the \cite{Ribas2015ProtoplanetaryPopulations} classification. Our model is therefore probing the regime where disk lifetime is expected to be the limiting factor giant planet formation. }

This can be explained in three main ways.

Firstly, that our model is an idealised system - a wholly smooth disk with only a single planet being formed in each iteration of our simulation.  Thus, there are no rings, gaps, discontinuities, or other growing planets competing for material.  Additionally, we are working under the assumption that giant planets are always formed via core accretion and disregard the possibility of disk fragmentation.  However, \cite{Cadman2020FragmentationStars} found that disks around stars more massive than $2 \Ms$ are more susceptible to gravitational instability and disk fragmentation.

Secondly, while we implement PMS stellar evolution, as detailed in Section \ref{sec:pms_ev}, we solely model photoevaporation via X-rays and do not include a shift in the dominant photoevaporation mechanism to FUV, corresponding to convective stars becoming radiative. Stars $\lesssim 2.5 \Ms$ mainly disperse via X-ray photoevaporation, but stars between $1.7{-}2.5 \Ms$ do become radiative and therefore hot enough for FUV to dominate on a timescale of around $5{-8}$ Myr that decreases with increasing stellar mass \citep{Kunitomo2021PhotoevaporativeStars}.  Stars more massive than $3 \Ms$ are hot enough to have a radiative core from the beginning \citep{Kunitomo2021PhotoevaporativeStars}. \cite{Kunitomo2021PhotoevaporativeStars} found that FUV is the main mechanism behind disk dispersal in stars more massive than $3 \Ms$.  Depending on how efficient FUV photoevaporation is at dispersing dust and gas in the disk, this could affect how many giant planets are able to form before limited material prohibits it.  Consequently, this could decrease the number of giants found in our model when $M_{\star} > 2.5 \Ms$.  In the context of the observational results, this might indicate that FUV is effective at removing material to cause such a low planet rate around stars more massive than $3 \Ms$.

Thirdly, our results may also reflect the age bias in observational accretion rates in Figure \ref{fig:accage}. As discussed in Section \ref{sec:acc}, %the observed accretion rates are when the star is anywhere from $1{-}14$ Myr for the \cite{Wichittanakom2020TheStars} data in our stellar mass range of interest.  Hence,
we may be underestimating the accretion rate of these stars by using them as the initial accretion rate assumption for our W20 model at 0.1 Myr.  For stars less massive than $1.25 \Ms$ in Figure \ref{fig:hist} (left), it gives a fairly good prediction for the rise of the giant planet occurrence rate within this stellar mass range.  But as detailed above, the 'peak' of our model distribution spans $M_{\star}{=}1.8{-}2.15 \Ms$, shifted slightly towards more massive stars than the observed peak.  Thus, it may be prudent for us to make our own estimation for this age discrepancy (J26).

To summarise, our W20 model gives a good approximation of the overall rise, peak, and fall of the giant planet occurrence rate as a function of stellar mass.  The overestimation of the planet rate can likely be attributed to the physical limitations of our model discussed above.  The broadness of the peak and the slow drop off for stars more massive than $2.5 \Ms$ may be due to the age bias present within our accretion rate estimations (Figure \ref{fig:accage}).  This leads us to explore our own estimation for the accretion rate, J26.  

Examining our J26 occurrence rate as a function of stellar mass in Figure \ref{fig:hist} (left), we can see that it provides a better fit to the observed and corrected occurrence rate than our W20 model. Specifically, we see a sharper rise in the planet rate up to $1.5 \Ms$ and a steeper decline for stars more massive than $2 \Ms$.  Now, our model rates lie - for the most part - within the limits of the corrected occurrence rate.  %Once again, this follows that the giant planet rate as a function of stellar mass is very sensitive to the initial accretion rate  

Adopting stellar-mass dependent accretion rates that account for the age bias in the observed accretion rate data (Figure \ref{fig:accage}) has narrowed and shifted the peak of the giant planet rate in our J26 model to better reflect the observational results.  The decline in the giant planet rate now falls off more quickly for stars more massive than $1.5 \Ms$ due to the higher accretion rates aiding in fast disk evolution and thus reducing the number of successful giant planets.  At the low end of our stellar mass range, there is a balance to be found: the accretion rate must be high enough to drive efficient pebble and gas accretion, yet not so high as to disperse the disk before a giant planet can form.  Observationally, this balance appears to be optimised around 1.7 M$_{\odot}$ \citep{Wolthoff2022PreciseSurveys}.
%Concurrently, at the low end of our stellar mass range, the balance for an 'ideal' accretion rate for giant planet formation \textbf{must be found} - high enough to make planetary pebble and gas accretion efficient while maintaining a sufficiently long disk lifetime - \textbf{as derived} observationally $\sim 1.7 \Ms$,  \cite{Wolthoff2022PreciseSurveys}. 

To summarise, our inclusion of stellar-mass dependent parameters in our models in this study has allowed us to recover giant planet occurrence rates that are compatible with observations.  We can directly compare this to results of our previous work \citep{Johnston2024FormationStars}, in which we were able to recover the declining $\eta_{\textrm{J}}$ behaviour for stars more massive than 2 M$_{\odot}$.  Without
the $M_{\star}$-dependence of the accretion rate, we could not capture the increasing $\eta_{\textrm{J}}$ rates for stars between $1{-}1.7$ M$_{\odot}$.  

%\subsubsection{Metallicity}

Now, when binned as a function of metallicity in Figure \ref{fig:hist} (right), we can see that the planet rate in both of our models (W20 and J26) increases with increasing metallicity up to 0.3 [dex].  Beyond this, we especially suffer from small number statistics which accounts for the low rates at this limit.  The J26 model produces a higher rate of giant planets for larger metallicities than our W20 model, likely due to the overall higher accretion rates in the J26 model which in turn helps aid the rate of pebble accretion. 

Within our models, metallicity is represented by the dust-to-gas ratio.  A larger metallicity means a higher ratio of solids to gas within our disks which enhances the pebble growth.  This allows the planetary embryos to grow more massive at a faster rate, meaning that a higher proportion of embryos can surpass isolation mass at an early enough stage to accrete enough gaseous material to become a giant planet.  This results in an increased giant planet occurrence rate for higher metallicities, as seen in Figure \ref{fig:hist} (right).  Overall, both of our models are in agreement with previous observational studies that found a positive planet-metallicity correlation \citep{Hekker2007PreciseParameters}.   %\citep{Wittenmyer2020CoolSearch,Wolthoff2022PreciseSurveys}.
Furthermore, several observational studies have also found that disks live longer around stars with higher levels of metallicity [Fe/H] \citep{Ercolano2010MetallicityLifetimes, Yasui2010ShortEnvironments, Nakatani2018RadiationDependence}.  The inclusion of this observed trend in our models may lead to a steeper occurrence rate and improve the model fit to the observations (Figure \ref{fig:hist}, right).  However, these results come from studies of low-mass stars and metallicities well below most of our sample; whether the trend holds for intermediate-mass stars is untested.

\section{Giant planet population properties}
\label{sec:prop} 
In this section, we explore the properties of the giant planets formed in our population synthesis models W20 and J26.  Namely: the final planetary masses, and the times and locations where these giant planets accumulate the bulk of their mass through runaway gas accretion.  These locations present starting points for subsequent inward migration, as long as the gas disc is sustained \citep{Kanagawa2018RadialPlanet}. We investigate how these properties relate to the stellar mass.

\subsection{Location and time of runaway gas accretion}
\label{sec:run}

\begin{figure}
\includegraphics[width=\linewidth]
{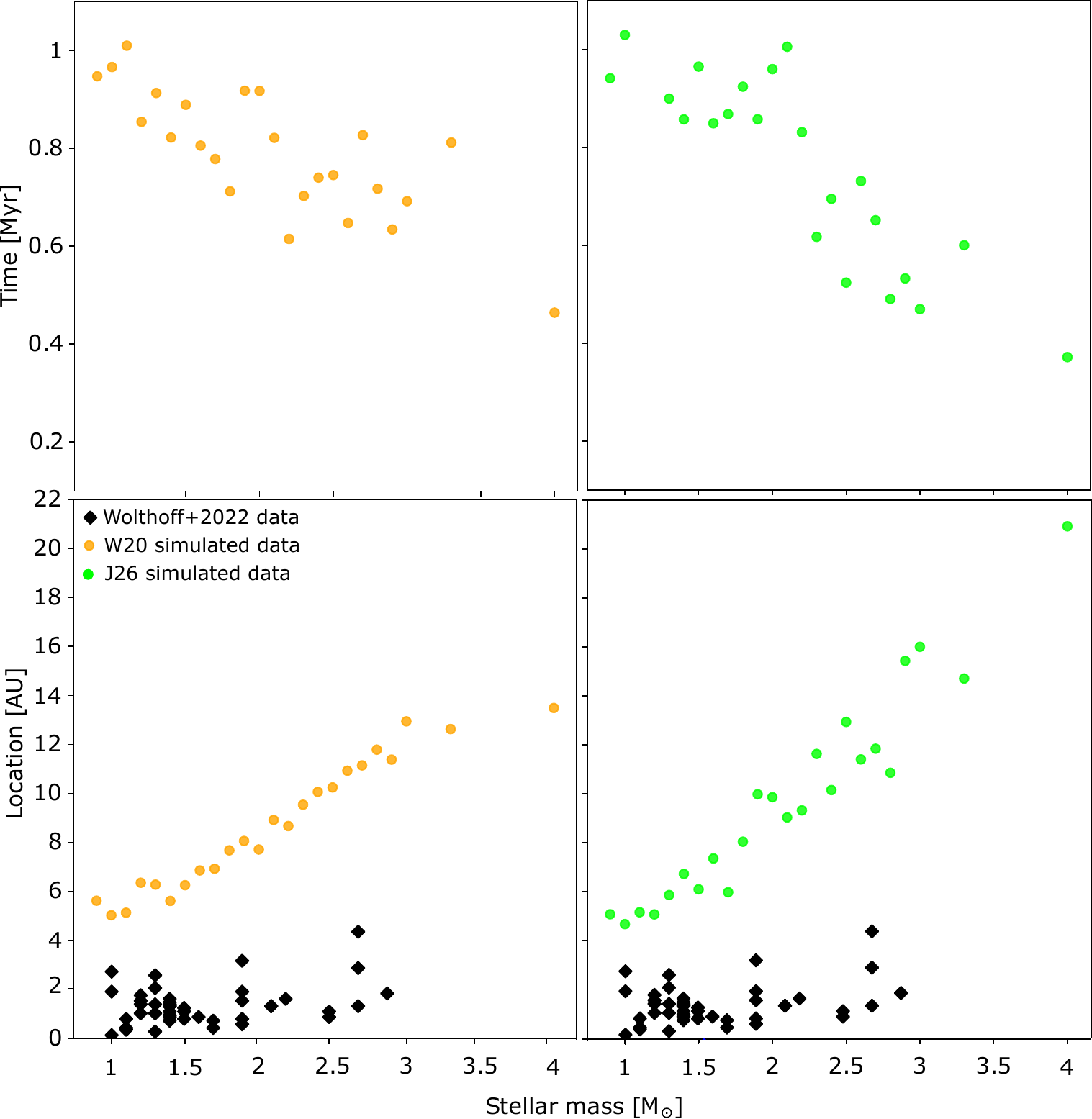}%{p2_giantprop_mstar.pdf}%{giantprop_mstar.pdf}
    \caption{Mean giant planet properties as a correlation of M$_{\star}$ of our planet population synthesis models W20 (orange) and J26 (lime).  The black diamonds are the 37 confirmed giant planets from the \citep{Wolthoff2022PreciseSurveys} sample.  Time and location refer to when and where our synthesised giant planets accumulate the bulk of their mass to surpass $300$ M$_{\oplus}$.}
    \label{fig:gprop_mstar}
\end{figure}

Figure \ref{fig:gprop_mstar} shows the mean times and locations where our giant planet population accumulates the bulk of its mass during the runaway gas accretion phase as a function of stellar mass.  The black diamonds are the 37 confirmed giant planets from the \cite{Wolthoff2022PreciseSurveys} sample - described in Section \ref{sec:sample}.  For both our W20 (orange) and J26 (lime) synthetic planet populations, clear trends are visible.  

Giant planets form earlier around more massive stars.  At $1 \Ms$, it takes around 1 Myr to grow a giant planet on average and this mean timescale has reduced to just 0.5 Myr by $4 \Ms$.  This trend follows directly from the stellar-mass dependent accretion rates discussed above.  High accretion rates - coupled with shortened disk lifetime and an enhanced rate of pebble accretion - mean that giant planets in such systems must form rapidly in order to exist at all.  The increasing $\dot{M}_{\rm g}$ with stellar mass corresponds to a higher solid accretion rate onto the planetary core (Equation \ref{eq:peb_acc}). Therefore, even though pebble isolation mass (Equation \ref{eq:m_iso}) scales with stellar mass, some embryos are able to accumulate enough solid mass very quickly when the gas and solid accretion rates are at their highest.  %The aforementioned trend of the high accretion rate reducing the disk lifetime around more massive stars means that there is a limited window for giant planet formation to proceed at early times at large orbital radii before the disk is dispersed.  
%This may have important implications for the wide-orbit giant planet rate which is also thought to peak around intermediate-mass stars \citep{Wagner2019OnPlanets}. } 
 
This is reflected further in the giant planet location plot of Figure \ref{fig:gprop_mstar}: the orbital location where giant planets accumulate the bulk of their mass increases with stellar mass, reaching $\gtrsim 10$ au for stars more massive than $\sim 2.5$ M$_{\odot}$.  Around low-mass stars the disk persists for longer, allowing giant planets to grow at a slower pace and complete runaway gas accretion within $\lesssim 5$ au.  Around more massive stars, the rapidly dispersing disk means that runaway gas accretion must occur at larger orbital radii before the available gas is removed.

A critical point regarding Figure \ref{fig:gprop_mstar} is that the simulated locations are \textit{not} the final orbital positions - they are the radii at which runaway gas accretion occurs, \textit{i.e.} the starting points for subsequent migration.  The observed giant planets (black diamonds) are all found within $\lesssim 5$ au, consistent with their having undergone substantial inward migration after the runaway gas accretion phase.  Indeed, the large offset between the simulated accretion locations ($5{-}20$ au) and the observed orbital radii ($\lesssim 5$ au) implies that significant inward migration is expected in all cases.  This is physically plausible: as long as the disk retains sufficient gas after the runaway accretion phase, migration will drive the planet inward on a timescale comparable to the local viscous timescale \citep{Kanagawa2018RadialPlanet}.  Since our model forms giant planets at early times $\lesssim 1$ Myr, upper panels of Figure \ref{fig:gprop_mstar}), there is substantial disk material remaining to facilitate this inward migration.  %This also provides a natural explanation for why our occurrence rates in Figure \ref{fig:hist} can match observations despite the apparent discrepancy in locations: the simulated planets are not stranded at their formation radii, and migration brings them into the detectable range of RV surveys.

That said, this migration is not guaranteed to be complete for all simulated planets, nor is it uniform.  Planets forming at the largest radii - particularly those around stars more massive than $\sim 2.5$ M$_{\odot}$ - may not migrate sufficiently far inward within the remaining disk lifetime to be detectable by RV surveys, which are generally sensitive to planets within $\lesssim 5$ au.  This suggests that some fraction of our simulated giant planets at large radii around massive stars would  \textit{not} be counted in the observed sample, which could contribute to the marginal overproduction of giant planets in our W20 model at high stellar masses in Figure \ref{fig:hist}.  A quantitative treatment of post-formation migration and its detectability is beyond the scope of this work, but represents an important avenue for future modelling. However, an important impediment is presented by engulfment of the innermost planets at later stages of stellar evolution \citep{Kunitomo2011PlanetGiants}.

We note that the gap between the simulated accretion locations (coloured circles) and the observed orbital radii (black diamonds) in the lower panels of Figure \ref{fig:gprop_mstar} can itself be interpreted as an observationally interesting parameter space. The region between $\sim 5{-}15$ au around intermediate-mass stars ($1.5{-}3$ M$_{\odot}$) corresponds to where giant planets in our model are born, and where some fraction of them may remain if migration is halted early.  For example, by the dissipation of the disk, or by trapping at a local pressure maximum \citep{Kanagawa2018RadialPlanet}.  This wide-orbit population would be largely inaccessible to current RV surveys, and the brightness of intermediate-mass host stars renders direct imaging at these separations extremely challenging due to the unfavourable contrast ratio \citep{Nielsen2019TheAU, Wagner2019OnPlanets}.  Astrometric surveys - in particularly those enabled by \textit{Gaia} and its successors - are therefore the most promising avenue for probing this parameter space and constraining what fraction of giant planets remain on wide orbits rather than migrating inward.
%but represents a promising target for direct imaging and astrometric searches.

%The disk - which dissipates from the inside-out - simply does not live for long enough when $\textrm{M}_{\star} \geq 3 \Ms$.  Thus, runaway gas accretion must occur at larger orbital radii ($\geq 10$ au).  The large orbital radii at which giant planets around more massive stars undergo runaway gas accretion has implications for their atmospheric composition.

%This has important consequences in terms of giant planet composition.  
The large orbital radii at which giant planets around more massive stars undergo runaway gas accretion also has implications for their atmospheric composition.  At such distances, H$_{2}$O and CO$_{2}$ are locked in the ice phase, while CO remains in the gas phase.  Any gas accreted in such a regime would be CO-rich, potentially producing a super-stellar C/O ratio in the planetary atmosphere.   In hot ($\rm T \gtrsim 1200$ K) photospheres, the H$_{2}$O abundance is sensitive to both the O/H and C/O ratios \citep{Welbanks2019K}.  A super-stellar C/O ratio can suppress the H$_{2}$O abundance and produce giant planets that are metal-rich but oxygen-poor \citep{Madhusudhan2014H2OJupiters, Welbanks2019K}.

Figure \ref{fig:gprop_mstar} includes only the mean-averaged values per stellar mass bin, and we note that small-number statistics affect the robustness of these averages at both ends of the stellar mass range, where our model giant planet rates fall below $5\%$ (Figure \ref{fig:hist}, left).  Additional observational detections at these limits would help constrain these trends. Nevertheless, since core accretion is thought to be the dominant giant planet formation mechanism \citep{Pollack1996FormationGas}, our model could be extended to incorporate disk structures such as rings and pressure bumps that could locally halt migration and retain planets on wider orbits. Additionally, a more complete treatment of photoevaporation across the full stellar mass range would be beneficial.

Overall, the accretion rate is a key parameter in determining the distribution of giant planets. The assumption of high accretion rates in the W20 and J26 models produces results consistent with current observational trends, supporting the hypothesis that (i) giant planet formation is dictated by the initial accretion rate and (ii) the orbital location of giant planets is largely dependent on the birth conditions of the natal disk.   %description of the giant planet formation mechanism in these models.

%\begin{figure}
%\includegraphics[width=\linewidth]{orbital_radius_white.pdf}
%    \caption{Comparison between the giant planet distribution predicted by the models J26 (top) and W20 (bottom) and observational data \citep{Wolthoff2022PreciseSurveys}. The heatmap shows %the kernel density estimate of 
%    the density of simulated giant planets \textbf{in the stellar mass vs. orbital radius (left) plane and metallicity vs. orbital radius (right) plane}.  White points represent observed exoplanets.}
%    \label{fig:obital-radius}
%\end{figure}

\subsection{Illustrative final planet masses and locations}

Sections \ref{sec:results} and \ref{sec:run} concern whether and when giant planets form. Here, we ask where they end up, which depends on processes that are poorly constrained or absent from our model. We follow growth and migration only until disk dispersal, in a smooth disk hosting one planet. Final radii and hot-Jupiter fraction depend on:

\begin{enumerate}
    \item The inner disk edge, which we set by magnetospheric truncation although strong fields are found in only a small fraction of Herbig Ae/Be stars;
    \item The timescale and mode of inner disk clearing around intermediate-mass stars, which are observationally unclear;
    \item Interactions with pre-existing inner planets and disk substructure, neither of which is modelled;
    \item Post-formation dynamics, which can change orbits dramatically.
\end{enumerate}

Moreover, the observed comparison sample consists of evolved stars.  \cite{Kunitomo2011PlanetGiants} show that the survival limit against engulfment falls steeply with mass, from $\sim 1$ au at 1.8 M$_{\odot}$ to $\lesssim 0.4$ au above 2 M$_{\odot}$, so a lack of planets at small radii cannot be read directly as primordial paucity.  Future surveys of young, intermediate-mass stars will likely measure the occurrence at small separations before engulfment. We regard the results below as illustrative and draw no conclusions from them.

%\textbf{We now compare the final locations and masses of our simulated giant planets with the observed giant planet population from the \cite{Wolthoff2022PreciseSurveys} sample to assess whether our models can match final planet formation outcomes. We define giant planets as those with final masses in the range of $0.8{-}24$ M$_{\textrm{Jup}}$, following \cite{Wolthoff2022PreciseSurveys}, and hot Jupiters as those with orbital separations $<0.1$ au (corresponding to periods $P\lesssim10$ days around solar-type stars), following \cite{Johnston2024FormationStars}.  We calculate planet growth and migration until disk dispersal, as in \citet{Johnston2024FormationStars, Johnston2025TheFormation}.}

\begin{figure}
    \includegraphics[width=\linewidth]{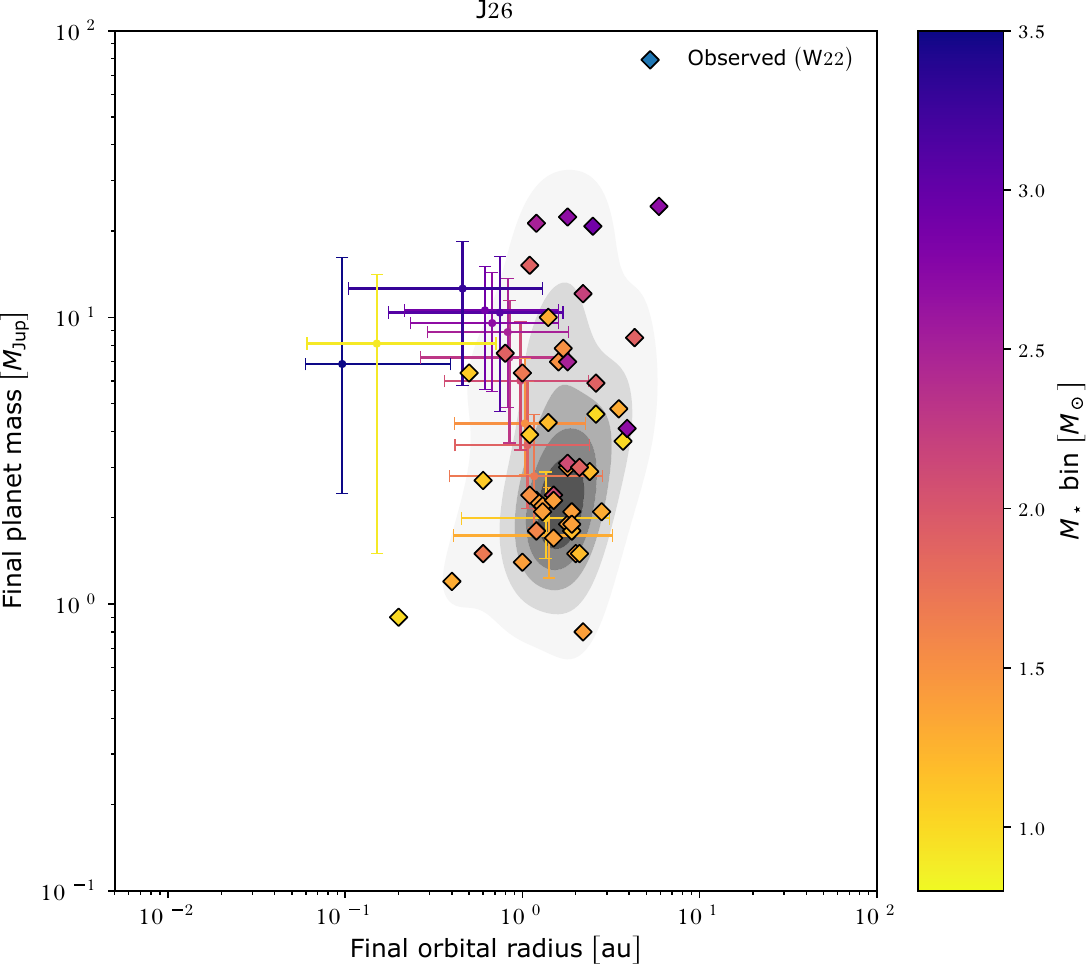}
    
\caption{Mean final giant planet properties of our planet population synthesis model (J26) binned by stellar mass. The error bars show the range of parameter space that we recover for our simulations in that stellar mass bin. The diamonds are the confirmed giant planets from the \citep{Wolthoff2022PreciseSurveys} sample.  
The colour refers to the mass of the host star. The filled contours represent kernel density estimations using a Gaussian kernel and a bandwidth chosen following Scott's rule \citep{Scott2015MultivariateVisualization}. }
\label{fig:gprop_fin}
\end{figure}

Figure \ref{fig:gprop_fin} shows the final orbital locations and mass of the J26 planet population as a function of stellar mass.  The observed systems are shown as diamonds colour-coded by stellar mass, while the simulated systems are shown as circles with the same colour scale.  The filled contours represent a Gaussian-kernel density estimate of the observed massive planet parameter space.  Most of the observed systems occupy the region between approximately $1$ and $3$ au and have reported masses of roughly $1{-}10$ M$_{\textrm{Jup}}$.  The observed masses are $M\sin i$ values and should therefore not be interpreted as exact planetary masses. 

The simulated J26 planets occupy the same broad region as the observations.  However, this figure alone does not provide a quantitative goodness-of-fit: the observed sample is small, subject to detection and selection effects, and has projected planetary masses. Moreover, the observed sample does not show a statistically robust monotonic relation between host-star mass and orbital radius. The apparent agreement should therefore be regarded as qualitative. %after migration, whereas the upper panels of Figure \ref{fig:gprop_mstar} show the locations at which runaway gas accretion occurs. The difference between these two locations represents the subsequent inward migration of planets.}

%\textbf{The observed systems overlap the broad mass--radius region occupied by the J26 planets. However, this figure alone does not provide a quantitative goodness-of-fit the observed sample is small, subject to detection and selection effects, and has projected planetary masses. Moreover, the observed sample does not show a statistically robust monotonic relation between host-star mass and orbital radius. The apparent agreement should therefore be regarded as qualitative. }

%\textbf{The final radii in the J26 model shift to shorter distances with increasing stellar mass. In the $1{-}1.4$ M$_{\odot}$ bins, the median final orbital radius is approximately $1.3{-}1.4$ au, while in the $2.6-3.4$ M$_{\odot}$ bins, it falls to roughly $0.6{-}0.8$ au. This trend is accompanied by substantial scatter, as evidenced by the error bars in Figure \ref{fig:gprop_fin}.  Furthermore, the highest-mass bins are limited by modest sample sizes. Thus, the model tentatively indicates a broad decline in final giant-planet orbital radius with host mass.}

With that caveat in mind, we report the median final orbital radii for completeness: approximately $1.3{-}1.4$ au in the $1{-}1.4$ M$_{\odot}$ bins and $0.6{-}0.8$ au in the $2.6{-}3.4$ M$_{\odot}$ bins. However, the scatter is substantial (Figure \ref{fig:gprop_fin}), and the highest-mass bins are limited by modest sample sizes. We therefore do not interpret these medians as evidence for a stellar-mass dependence of the final orbital radius.

In our earlier work \citep{Johnston2024FormationStars}, giant planets formed at several au and underwent substantial inward migration. At 1 M$_{\odot}$, that work recovered a hot Jupiter occurrence rate of $\sim 0.7{-}0.8 \%$. In the updated models presented here, the hot Jupiter rate across the whole J26 population is $\sim 0.95\%$. We do not interpret this difference, as it depends on the assumed inner edge and the uncertainties listed above. 

%the final orbital radii decline from a median of $\sim 1.3{-}1.4$ au at 1 M$_{\odot}$ to $\sim 0.6{-}0.8$ au at 3 M$_{\odot}$ (Figure \ref{fig:gprop_fin}).}

The disk inner edge in our model (Equation \ref{eq:r_in}) acts as the location where inward migration and solid accretion are halted.  We assume that this edge is set by stellar magnetospheric torques \citep{Lin1996OrbitalLocation}.  However, strong magnetic fields are detected in only a fraction of Herbig Ae/Be stars (\textit{e.g.} \cite{Alecian2013AMeasurements}).  If the inner disk edge is absent or located elsewhere, planets could migrate farther inward or be lost to the host star. Consequently, this would decrease our hot-Jupiter occurrence rate for intermediate-mass stars only.  Additionally, our assumption of a smooth, radially symmetric disk neglects structures such as rings and gaps that could halt migration at large radii \citep{Kanagawa2018RadialPlanet}.

Finally, the J26 simulations show a broad decline in final giant planet orbital radius with increasing stellar mass, while the observed sample shows no comparable trend. The comparison is limited by the small observed sample; uncertainty in the inner disk edge; and the absence of substructure(s) in our model. We therefore do not interpret the model trend as a prediction, though future astrometric and direct-imaging surveys may be able to test whether such a trend exists.

\section{Summary and Conclusions}
\label{sec:conc}

We have presented a planet population synthesis study of giant planet formation across a range of stellar masses and metallicities, using the pebble-driven core accretion model of \cite{Liu2019Super-EarthMasses}, as adapted in \cite{Johnston2024FormationStars, Johnston2025TheFormation}.  %We conduct planet population synthesis using the disk model from \cite{Liu2019Super-EarthMasses, Johnston2024FormationStars} and the stellar parameters from the homogenised stellar sample of three combined RV surveys from \cite{Wolthoff2022PreciseSurveys}.  
We choose the initial conditions of planet formation such that they are close to current observational trends.  We investigate three different best fits for the observational accretion rate; (i) M17 for the low-mass stars in \cite{Manara2017X-ShooterStars}; (ii) W20 for the intermediate-mass stars in \cite{Wichittanakom2020TheStars}; and J26 as our own estimation to account for the wide range of ages present in the \cite{Wichittanakom2020TheStars} sample.

We find that by adopting observational stellar mass-dependent accretion rates (J26, Figure \ref{fig:accage}); disk lifetimes (Equation \ref{eq:pho}); and PMS stellar evolution (Section \ref{sec:pms_ev}), we can produce giant planet rates as functions of stellar mass and metallicity that corresponds to the observed distribution in \cite{Wolthoff2022PreciseSurveys} (Figure \ref{fig:hist}).  Furthermore, it is evident that our giant planet population models are highly sensitive to the initial accretion rate (Figures \ref{fig:heat} \& \ref{fig:hist}).  When we adopt the observed low accretion rates around less massive stars, the low accretion rates suppress the efficient pebble growth and gas accretion required to form giant planets.  Conversely, when we adopt the observed high accretion rates around more massive stars, the disk is dissipated at a faster rate which results in the embryos being quickly swept to the edge of the inner disk and hence preventing the growth of giant planets. This gives rise to the peak of occurrence rate around 1.75~M$_\odot$, as derived observationally.  

As a consequence of adopting this stellar-mass dependent accretion rate, we find that giant planets around more massive stars form at earlier times than around less massive stars, which counteracts the fast-acting disk dissipation (Figure \ref{fig:gprop_mstar}).  Additionally, these giant planets around more massive stars tend to accrete the bulk of their mass at greater distances from their host stars, allowing sufficient time for the core to acquire enough solid mass needed for the runaway gas accretion, before reaching the edge of the inner disk (Figure \ref{fig:gprop_mstar}). 

Our results from Figure \ref{fig:gprop_mstar} indicate that giant planets can form at a range of radii, including well beyond the $\rm H_{2} \rm O$ snowline, without requiring icy grain enhancement of solid accretion.  In fact, many of our successful giant planets are born well beyond the $\rm H_{2} \rm O$ snowline.  This notably changes the classical view of giant planet formation via core accretion that often relies on the enhanced sticking efficiency of icy grains beyond the snowline to accelerate planetesimal and core growth.  Instead, the stellar-mass dependent accretion rate alone appears sufficient to drive efficient core growth across a range of disk radii, in the context of pebble accretion driven planet formation models.

%\textbf{Lastly, we compare how well the orbital distances of our model giant planet reflect the 37 giant planets from the \cite{Wolthoff2022PreciseSurveys} sample as a function of both stellar mass and metallicity (Figure \ref{fig:obital-radius}.  

We also find that both our J26 and W20 models provide a reasonable agreement with the observed giant planet properties.  This reaffirms our findings that accretion rate is the most important parameter in dictating whether giant planets can form in a system.  Additionally, it provides an exciting avenue for future work: if we can successfully model the earliest stage of planet formation (\textit{i.e.} giant planets), we can explore  terrestrial planet formation from the remaining disk material and investigate the resulting planetary system architecture through dynamical modelling.  %This could potentially lead to us identifying promising candidates for Solar System analogues from an initial giant planet detection. \textbf{OP: This last sentence is a bit confusing to me...} 

Moreover, future exoplanet observations - not only using radial velocity but astrometric methods such as those enabled by \textit{Gaia} and forthcoming missions - have the potential to substantially improve our understanding of giant planet occurrence rates in the parameter space most relevant to giant planet formation and migration.  As illustrated in Figure \ref{fig:gprop_mstar}, the region between the simulated runaway accretion locations ($5{-}20$ au) and the currently observed orbital radii ($\lesssim 5$ au) represents a largely unexplored parameter space for intermediate-mass host stars.  Astrometric surveys are particularly well-suited to probing this regime and would provide direct observational constraints on whether giant planets formed at large orbital radii subsequently migrate inward, remain on wide orbits, or are dispersed entirely.  Each outcome carrying distinct implications for the giant planet occurrence rates discussed in this work.

Finally, our current model is limited by several assumptions which could explain our marginal overproduction of giant planets in Figure \ref{fig:hist}.  Specifically, our assumption of a  radially smooth disk neglects structures such as rings and gaps that could locally inhibit migration.   In reality, the planets could be stopped at larger orbits by processes not included in our model.  Furthermore, planets may have been removed in some stars due to engulfment during later stages of evolution.  In future work, we hope to introduce some rings or gaps that would naturally inhibit planetary growth and migration.  In conjunction with this, adopting a disk chemistry model would also allow us to find any chemical composition distinctions that may exist in the giant planet population, either as a function of stellar mass or metallicity.

\section*{Acknowledgements}

  This work was undertaken on ARC4, part of the High Performance Computing facilities at the University of Leeds, UK.  We thank both Richard Booth and Daniela Iglesias for their helpful inputs. We thank the reviewer for the helpful report.  HFJ was supported by a Royal Society Grant (RF-ERE-221025) and the Science and Technology Facilities Council (Project Reference: 2441776). OP acknowledges support from the Science and Technology Facilities Council, grant numbers ST/T000287/1, and ST/X001016/1.  BL is supported by National Key R\&D Program of China (2024YFA1611803), National Natural Science Foundation of China (Nos. 12222303 and 12173035), the start-up grant of the Bairen program from Zhejiang University.

%%%%%%%%%%%%%%%%%%%%%%%%%%%%%%%%%%%%%%%%%%%%%%%%%%
\section*{Data Availability}

The data underlying this article will be shared on reasonable requests to the corresponding author.

%%%%%%%%%%%%%%%%%%%% REFERENCES %%%%%%%%%%%%%%%%%%

% The best way to enter references is to use BibTeX:

\bibliographystyle{mnras}
\bibliography{references} % if your bibtex file is called example.bib

% Alternatively you could enter them by hand, like this:
% This method is tedious and prone to error if you have lots of references
%\begin{thebibliography}{99}
%\bibitem[\protect\citeauthoryear{Author}{2012}]{Author2012}
%Author A.~N., 2013, Journal of Improbable Astronomy, 1, 1
%\bibitem[\protect\citeauthoryear{Others}{2013}]{Others2013}
%Others S., 2012, Journal of Interesting Stuff, 17, 198
%\end{thebibliography}

%%%%%%%%%%%%%%%%%%%%%%%%%%%%%%%%%%%%%%%%%%%%%%%%%%

%%%%%%%%%%%%%%%%% APPENDICES %%%%%%%%%%%%%%%%%%%%%

%\appendix
%\section{Additional planet population synthesis}

%%%%%%%%%%%%%%%%%%%%%%%%%%%%%%%%%%%%%%%%%%%%%%%%%%

% Don't change these lines
\bsp	% typesetting comment
\label{lastpage}
\end{document}